\documentclass[printer]{tJOT2e_edited}

\usepackage{rotating}
\newcommand{\sref}[1]{section~\ref{#1}}
\newcommand{\eref}[1]{equation (\ref{#1})}
\newcommand{\fref}[1]{figure~\ref{#1}}
\newcommand{\tref}[1]{table~\ref{#1}}

\newcommand{\Fref}[1]{Figure~\ref{#1}}
\newcommand{\Tref}[1]{Table~\ref{#1}}

\newcommand{\opensquare}{\mbox{$\rlap{$\sqcap$}\sqcup$}}

\newcommand{\dashed}{\protect\mbox{-~-~-~-}}
\newcommand{\broken}{\protect\mbox{--~--~--~--}}

\newcommand{\chain}{\protect\mbox{--- $\cdot$ ---}}

\newcommand{\full}{\protect\mbox{------}}

\newcommand{\etal}{{\it et al}.\,}

\begin{document}
\doi{}
\issn{}
\issnp{} \jvol{00} \jnum{00} \jyear{2014} \jmonth{May}

\markboth{T. Tsukahara et al.}{Turbulent channel flow at low Reynolds numbers}

\title{DNS of turbulent channel flow at very low Reynolds numbers}

\author{Takahiro TSUKAHARA$^{*}$\thanks{${~\,}^*$Corresponding author. Department of Mechanical Engineering, Tokyo University of Science, 2641 Yamazaki, Noda-shi, Chiba 278-8510, Japan.  E-mail: tsuka@rs.tus.ac.jp}, 
Yohji SEKI,
Hiroshi KAWAMURA,
Daisuke TOCHIO
\\
\thanks{\vspace{6pt}\newline\centerline{\tiny{ {\em Preprint at arXiv} 
\textcopyright 2014 Tsukahara Lab, Tokyo University of Science}}}
\affil{}}

\received{}

\maketitle

\begin{abstract}
Direct numerical simulations (DNS) of fully-developed turbulent channel flows for very low Reynolds numbers have been performed with a larger computational box sizes than those of existing DNS. 
The friction Reynolds number was decreased down to 60, where the friction Reynolds number is based on the friction velocity and the channel half width.
When the Reynolds number was decreased to 60 with small computational box size, the flow became laminar.
Using a large box, we found that a localized turbulence was observed to sustain in the form of periodic oblique band. 
This type of locally disordered flow is similar to a equilibrium turbulent puff in a transitional pipe flow.
Various turbulence statistics such as turbulence intensities, vorticity fluctuations and Reynolds stresses are provided. 
Especially, their near-wall asymptotic behavior and budget terms of turbulence kinetic energy were discussed with respect to the Reynolds-number dependence and an influence of the computational box size. 
Other detailed characteristics associated with the turbulence structures were also presented and discussed.
\end{abstract}

%%====================================== INTRODUCTION
\section{Introduction}
\label{sec:intro}

Low-Reynolds-number turbulent flow in a channel is of practical importance with respect to many engineering applications, such as heat exchange equipment. 
Laminar flows show much smaller drag, mixing and heat transfer than turbulent flows do.
Transition from a turbulent flow to a laminar flow --- so-called laminarization --- has been studied experimentally by a number of researchers, e.g., Narasimha \& Sreenivasan \cite{Narashimha79}, since the process of turbulent-laminar transition is also important in both the fields of industrial applications and fundamental flow physics.

Direct numerical simulation (DNS, hereafter) of a fully developed turbulent channel flow has been increasingly performed for higher Reynolds numbers with an aid of recent development of computers. 
DNS provides various information, such as velocity and pressure. 
Special attention has been paid to their near-wall asymptotic behavior and their derivatives at any time and point in an instantaneous field, which are extremely difficult to be measured in experiments. 

The first DNS of the fully developed turbulent channel flow was made by Kim \etal \cite{Kim87}. 
Their Reynolds number was $Re_\tau=180$, which is based on the channel half width $\delta$, the kinematic viscosity $\nu$ and the friction velocity $u_\tau=\sqrt{\tau_w/\rho}$, where $\tau_w$ is the statistically averaged wall shear stress and $\rho$ is the density.
Kuroda \etal \cite{Kuroda89} carried out the DNS for a slightly lower Reynolds number of $Re_\tau$=150. 
Kawamura and co-workers \cite{Kawamura98, Abe01, Abe04a, Abe04b} performed the DNS with respect to Reynolds number and Prandtl number dependences for $Re_\tau=180$--1020. 
As for low Reynolds numbers, several research groups carried out DNS or LES down to $Re_\tau=80$ in order to study turbulence control\cite{Bewley01, Chang02, Hogberg03}.
Iwamoto \etal \cite{Iwamoto02} have executed DNS for $Re_\tau=110$--650: their published results of $Re_\tau=110$ and 150 are also included in this paper for comparison. 
Iida \& Nagano \cite{Iida98} performed DNS for $Re_\tau=60$--100 to investigate flow fields with emphasis on the streamwise vortexes.
These studies employed rather small computational boxes because of limited calculation resources. 

A great deal of effort has also been devoted to experimental studies of the turbulent channel flow. 
Laufer \cite{Laufer51} first obtained the detailed turbulence statistics in the channel flow. 
Patel \& Head \cite{Patel68,Patel69} measured skin friction and mean velocity profiles in a range of $Re_{\rm m}=1,000$--10,000, which includes the transition from the laminar to the turbulent flow. 
The bulk Reynolds number $Re_{\rm m}$ is based on the bulk mean velocity $u_{\rm m}$ and the channel width.
Their measurement with a hot-wire showed that turbulent bursts occurred above $Re_{\rm m}=1,380$.
Later, Kreplin \& Eckelmann \cite{Eckelmann74,Kreplin79} made their experiments for low Reynolds numbers of $Re_{\rm c}=2,800$--3,850, based on the centreline velocity $u_{\rm c}$ and $\delta$. 
The measurements of the turbulent channel flow at $Re_{\rm c}=2,850$ and 3,220 were executed by Niederschulte \etal \cite{Niederschulte90}. 
Durst \& Kikura \cite{Durst95} carried out their experiments by means of laser-Doppler anemometry (LDA) at Reynolds number range of $Re_{\rm m}=2,500$--9,800, which corresponds to $Re_\tau=87$--293.

It should be emphasized that comparing of the existing experimental results \cite{Davies28,Kao70,Nishioka75} showed wide variation in the critical Reynolds numbers from $Re_{\rm c}=266$ to 8000.
This is because the transition in plane channel flows is sensitive to the background turbulence, e.g., inlet and initial conditions.
According to the linear instability analysis using the Orr-Sommerfeld/Squire equations, no exponentially growing solution is found below $Re_{\rm c}=5,772.22$ \cite{Orszag71}. 
Later, theoretical results of Orszag's group \cite{Orszag80a,Orszag80b} inferred a transitional Reynolds number of about 1,000, when finite-amplitude three-dimensional disturbances were considered.
Other researchers also indicated that there is a possibility of transient energy growth of disturbance at a Reynolds number as low as $Re_{\rm c}=1,000$ \cite{Kleiser91,Butler92,Reddy93}.

With the aid of developed supercomputing system, DNS of a turbulent channel flow at high Reynolds number has been carried out with a large computational domain, since much attention is paid to a large-scale motions (LSM) in the outer region\cite{Jimenez98, Liu01, Abe04a, Abe04b}.
For a lower Reynolds number, on the other hand, near-wall streaky structures are so elongated that their lengths exceed the usual computational box sizes. 
Thus DNS of a low Reynolds number flow also requires a larger box size to capture the near-wall streaky structures and the LSM. 

In the present work, DNS of a fully-developed turbulent channel flow has been carried out with  larger computational boxes than those of previous works. 
The purpose of this study is to obtain the turbulence statistics and structures of the turbulent channel flow at very low Reynolds numbers, cf. \tref{tab:re}.

\section{Numerical procedure}

%%====================================== TABLE: REYNOLDS NUMBER
\begin{table}[t]
 \tbl{Reynolds numbers of the present DNS and friction coefficient $C_f$. 
      Reynolds numbers: $Re_\tau$=$u_\tau\delta/\nu$, $Re_{\rm m}$=$u_{\rm m}2\delta/\nu$ 
      and $Re_{\rm c}$=$u_{\rm c}\delta/\nu$.
      Computational domain size: MB, medium box size; LB, large box size; 
      XL, extra-large box size, cf., \tref{tab:box}.
     }
 {\begin{tabular}{lcccccccc}\toprule
$Re_\tau$    & 180  & 150  & 110  & ~80  & ~80  & ~70  & ~64  & ~60$^{\rm a}$ \\\colrule
$Re_{\rm m}$ & 5730 & 4620 & 3290 & 2290 & 2320 & 2010 & 1860 & 1580$^{\rm b}$ \\
$Re_{\rm c}$ & 3360 & 2720 & 1960 & 1400 & 1430 & 1270 & 1200 & ~930$^{\rm b}$ \\
Box size     & MB   & MB   & MB   & MB   & XL   & LB   & LB   & LB \\
$u_{\rm c}/u_{\rm m}$& 1.17 & 1.18 & 1.19 & 1.22 & 1.24 & 1.26 & 1.29 & ----- \\
$C_f\times 10^3$     & 7.90 & 8.42 & 8.95 & 9.60 & 9.52 & 9.65 & 9.40 & ----- \\
  \botrule
  \end{tabular}}
  \tabnote{$^{\rm a}$ This case resulted in a laminarization.}
  \tabnote{$^{\rm b}$ A value estimated by equations (\ref{eq:emp1}) and (\ref{eq:emp2})}
 \label{tab:re}
 \vspace{-0.8em}
\end{table}
%%=============================================================

%%=========================================== CONFIGURATION
\begin{figure}[tbp]
 \begin{center}
  \epsfxsize=80.0mm
  \epsfbox{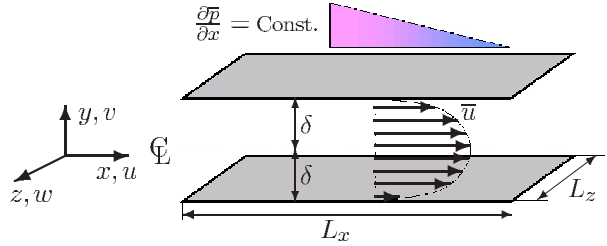}\\
 \caption{Channel configuration.}
 \label{fig:domain}
 \end{center}
\end{figure}
%%=============================================================

The mean flow is driven by the uniform pressure gradient: see \fref{fig:domain}. 
It is assumed to be fully developed in the streamwise ($x$-) and spanwise ($z$-) directions. 
The coordinates and the flow variables are normalized by $u_\tau$, $\nu$ and $\delta$. 
Periodic boundary conditions are imposed in the $x$- and $z$-directions and a non-slip condition is applied on the walls. 
The fundamental equations are the continuity and the Navier-Stokes equations:
\begin{equation}
 \frac{\partial u_i}{\partial x_i}=0,
 \label{eq:continuity}
\end{equation}
\vspace{-1em}
\begin{equation}
\frac{\partial u_i^*}{\partial t^*}
+{u_j^*}\frac{\partial u_i^*}{\partial x_j^*}
=-\frac{\partial p^*}{\partial x_i^*}
+\frac{1}{Re_\tau} \frac{\partial ^2 u_i^*}{\partial {x_j^*}^2},
 \label{eq:ns}
\end{equation}
\noindent{}where $u_i$, $t$ and $p$ are velocity vector, time and pressure, respectively.
Note that the superscript $*$ indicates a quantity normalized by $u_\tau$ and/or $\delta$.

For the spatial discretization, the finite difference method was adopted. 
The numerical scheme with the 4th-order accuracy was employed in the $x$- and $z$-directions, while the one with the 2nd-order is applied in the $y$-direction. 
Time advancement was executed by the semi-implicit scheme: the 2nd-order Crank-Nicolson method for the viscous terms on the $y$-direction and the 2nd-order Adams-Bashforth method for the other terms.

%%====================================== TABLE: DOMAIN SIZE
\begin{table}[btp]
 \tbl{Computational domain size; $L_i$, $N_i$ and $\Delta i$ are a box length, 
      a grid number and a spatial resolution of $i$-direction, respectively.}
 {\begin{tabular}{cccc}\toprule
Box size                  & MB          & LB           & XL           \\\colrule
$L_x\times L_y \times L_z$& $12.8\delta\times 2\delta \times  6.4\delta$ 
                          & $25.6\delta\times 2\delta \times 12.8\delta$ 
                          & $51.2\delta\times 2\delta \times 22.5\delta$ \\
$N_x\times N_y \times N_z$& $ 256 \times 128 \times 256$ 
                          & $ 256 \times 128 \times 256$ 
                          & $1024 \times  96 \times 512$ \\
$\Delta x^*, \Delta z^*$  & 0.05, 0.025 & 0.10, 0.05   & 0.05, 0.044  \\
$\Delta y_{\tiny{\textrm{min}}}^*$--$\Delta y_{\tiny{\textrm{max}}}^*$ & 0.0011--0.033 
                          & 0.0011--0.033 & 0.0014--0.045\\\botrule
  \end{tabular}}
 \label{tab:box}
\end{table}
%%=========================================================

In the present work, a series of DNS were made for $Re_\tau=60$--180.
\Tref{tab:re} summarizes the friction Reynolds numbers $Re_\tau$ and some mean-flow parameters.
The computational domain should capture at least a couple of near-wall low-speed streaks, and $L_i^*$ should be relatively large in a low-Reynolds-number DNS because of existence of an elongated streaky structure.
For the lower Reynolds numbers of $Re_\tau=60$--80, the larger boxes than those in the literature were adopted. 
The computational conditions are shown in \tref{tab:box}.
The stream- and spanwise lengths of the computational domain of XL were chosen as $L_x^*  \times L_z^*=51.2 \times 22.5$, which were approximately 4100 and 1800 wall units (at $Re_\tau=80$), respectively, whereas the characteristic sizes of the near-wall structure are $\lambda_x^+ \approx 1000$ and $\lambda_z^+ \approx 100$.
This was found to be large enough to capture the elongated near-wall streaky structure.

The non-uniform meshes were applied in the wall-normal direction.
Abe \etal \cite{Abe01} confirmed, \emph{a posteriori}, that the 2nd-order scheme retains high accuracy on a non-uniform mesh and acceptable results can be obtained when fine enough grids are adopted. 
The minimum wall-normal grid spacing was approximately equal to $0.3 \eta$ ($\eta$ is referred to as a local Kolmogorov scale) at $Re_\tau=180$.
In the present study, a coarser mesh ($N_y =96$) was used only for $Re_\tau=80$~(XL).
In this case, the grid spacings were $\Delta y^+=0.22$--$3.59$, which corresponded to $0.13\eta$--$1.25\eta$.
With the finer mesh of $N_y =128$, the maximum spacing (at the centreline of the channel) is also adjusted to keep the resolution, less than $1.6\eta$, for the other Reynolds numbers.
The grid resolutions of the present simulations are sufficiently fine to resolve the essential turbulent scales, and the reliable results can be achieved even with the 2nd-order scheme in the wall-normal direction.

In the present DNS, the pressure gradient was decreased stepwise down to an estimated level for an aimed Reynolds number.
A fully-developed flow field at a higher Reynolds number was successively used as the initial condition for a one-step lower Reynolds number, e.g., $Re_\tau=180\rightarrow$150, 150$\rightarrow$110, 110$\rightarrow$80 and so on.
Note that various statistical data were obtained after the flow had reached a statistical-steady state.

\section{Results and Discussion}

\subsection{Mean velocity profile}

%%-------------------------------------- U_MEAN
\begin{figure}[t]
 \centerline{\epsfxsize=120.0mm
  \epsfbox{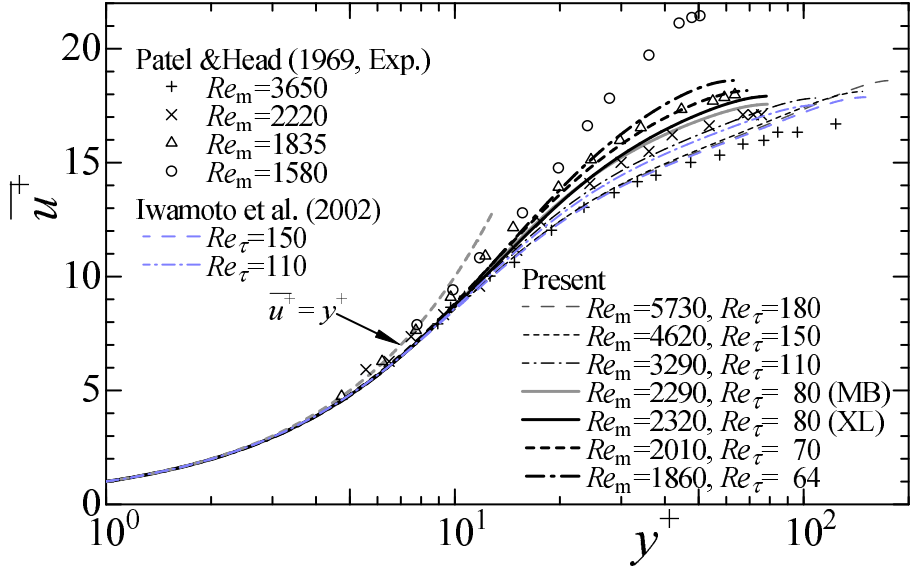}}
 \vspace{-0.0em}
 \caption{Mean velocity profiles in viscous wall units.}
 \label{fig:umean}
\end{figure}

\Fref{fig:umean} shows the wall-normal profile of the dimensionless mean velocity in the wall units, where quantities with the superscript of ${}^+$ indicate those normalized by the wall variables, e.g., $y^+ = yu_\tau/\nu$.
Statistics are denoted by an overbar of $\overline{(\,)}$, that are the spatial (in the horizontal directions) and temporal averaging.
The obtained data from the experiments of Patel \& Head \cite{Patel69} and the DNS by Iwamoto \etal \cite{Iwamoto02}, who used the spectral method are also shown for comparison.
The present results for $Re_\tau=110$--180 are in reasonable agreement with the existing DNS. 
For the lower Reynolds numbers of $Re_\tau\leq110$, the Reynolds-number dependence in the DNS data is also consistent with the trend of 
the mean velocity profiles measured by Patel \& Head \cite{Patel69}.
In these low Reynolds number flows, the mean velocity distributions do not indicate an evident logarithmic region. 
Thus, the von K\'arm\'an constant (not shown here) does not exhibit any constant range at all.
In the outer region ($y^+>10$), a significant Reynolds-number dependence is found when normalized by the inner variables. 

%%-------------------------------------- UM, UC
\begin{figure}[t]
 \centerline{\epsfxsize=120.0mm
 \epsfbox{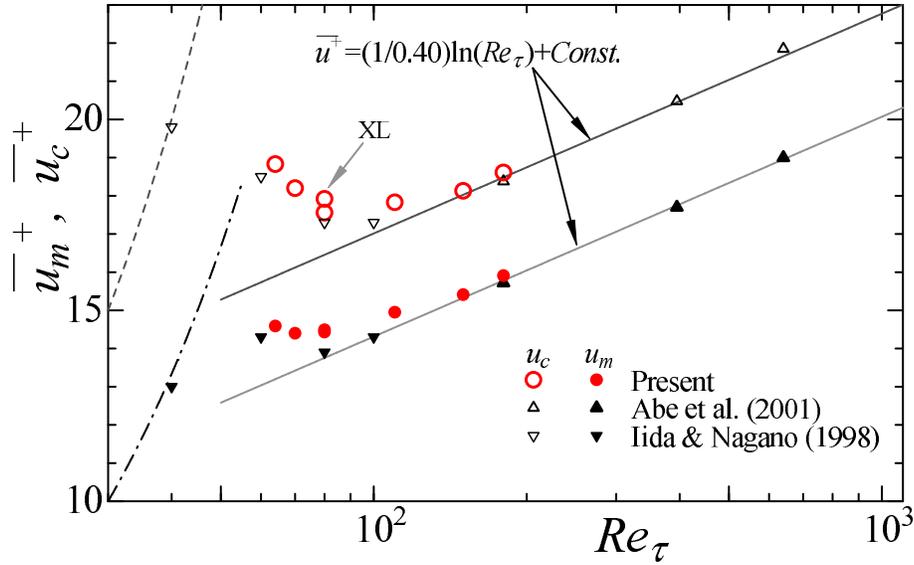}}
 \vspace{+0.2em}
 \caption{Reynolds number \emph{versus} centreline velocity $u_{\rm c}^+$ and 
          bulk mean velocity $u_{\rm m}^+$.
          Dotted lines (\broken, \chain) show the laminar flow relations of 
          $u_{\rm c}^+=Re_\tau/2$ and $u_{\rm m}^+=Re_\tau/3$, respectively.
          }
 \label{fig:umuc}
\end{figure}

The mean flow variables such as the bulk mean velocity $\overline{u}^+_{\rm m}$ and the mean centreline velocity $\overline{u}^+_{\rm c}$ are given in \fref{fig:umuc} for each Reynolds number.
It is interesting to note that the values of $\overline{u}^+_{\rm c}$ and $\overline{u}^+_{\rm m}$ increase with decreasing Reynolds number for $Re_\tau\le 80$. 
Both $\overline{u}^+_{\rm c}$ and $\overline{u}^+_{\rm m}$ are expected to approach gradually to laminar values.
Dean \cite{Dean78} found that the ratio between $u_{\rm c}$ and $u_{\rm m}$ was given by 
\begin{equation}
  \frac{u_{\rm c}}{u_{\rm m}}=1.28 Re_{\rm m}^{-0.0116}
  \label{eq:emp1}
\end{equation}
for a turbulent channel flow.
In the range of the present Reynolds numbers, the increasing rate of $u_{\rm c}/ u_{\rm m}$ (see \tref{tab:re}) with decreasing $Re_{\rm m}$ is larger than the one indicated by \eref{eq:emp1}.
The discrepancy between the obtained $u_{\rm c}/ u_{\rm m}$ and \eref{eq:emp1} reflects the low Reynolds-number effect.
If emphasis is placed on the data at $Re_\tau =80$ (MB and XL), one can see slight increases in $u_{\rm c}$, $u_{\rm m}$ and $u_{\rm c}/ u_{\rm m}$ with extending the box size since the quasi-laminar region locally appeared in the flow field of XL, as will be discussed in \sref{sec:structure}.

\subsection{Laminarization}
\label{sec:um}

%%-------------------------------------- CF
\begin{figure}[tbp]
 \centerline{\epsfxsize=120.0mm
 \epsfbox{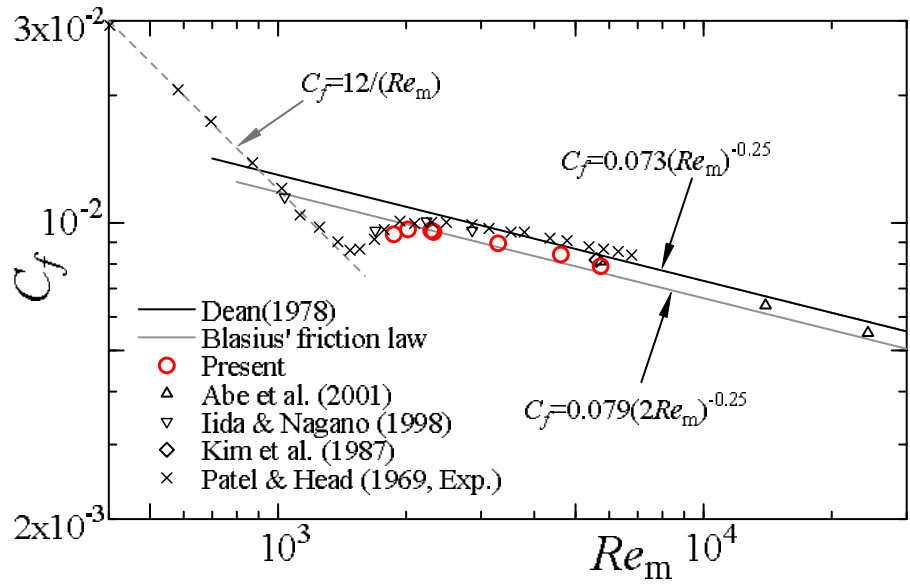}}
 \vspace{-0.0em}
 \caption{Variation with bulk Reynolds number of $C_f$.}
 \label{fig:cf}
\end{figure}

A skin friction coefficient in the transitional region is one of the most fundamental turbulence characteristics. 
A large number of experimental studies have been devoted to this issue, while it has not been examined yet through DNS owing to the lack of the low Reynolds number simulations.
\Fref{fig:cf} shows a variation of the skin friction coefficient $C_f$=$2\tau_w/(\rho\cdot u_{\rm m}^2)$ of the channel flow in comparison with the empirical correlations proposed by Blasius for a pipe flow, and by Dean \cite{Dean78} for the channel flow as, 
\begin{equation}
  C_f=0.073(Re_{\rm m})^{-0.25}.
  \label{eq:emp2}
\end{equation}
Both the measurement result and the numerical data from the previous DNS on the turbulent channel flows are also shown in \fref{fig:cf}. 

The present results are in good agreement with the reference data for $Re_{\rm m}>3,000$. 
It is worth noting that, for $Re_{\rm m}<3,000$, $C_f$ tends to be smaller than the empirical correlations and decreases for $Re_{\rm m}$ less than about 2,000.
This tendency agrees well with the data from the measurement of Patel \& Head \cite{Patel69} and the DNS of Iida \& Nagano \cite{Iida98}. 
The approximate value of 2,000 mentioned above roughly coincides with a critical Reynolds number, below which the flow became intermittently laminar and turbulent in the experiment by Patel \& Head \cite{Patel68,Patel69}.
They also showed that the ($C_f$ \emph{vs.} $Re_{\rm m}$)-curve, not shown in \fref{fig:cf}, remained unaltered for different entry conditions.
This implies that the decrease in $C_f$ would be a universal phenomenon independent of the inlet condition in the case of their experiment.
In the present investigation, there may be also inconsiderable differences between the values (of $C_f$ or $u_{\rm m}$) in the two flows of $Re_\tau=80$~(MB and XL), although an evidence of the effect of box size appears in the $u_{\rm c}$ as described above.

The second lowest Reynolds number ($Re_\tau=64$) of the present study, which corresponds to $Re_{\rm m}=1,860$ in the present DNS, still lies in the transitional flow regime. 
Further decrease in the Reynolds number below $Re_\tau=60$ resulted in a laminar flow field.
On the other hand, the DNS performed by Iida \& Nagano \cite{Iida98} revealed that the flow field  remains turbulent accompanied with a temporally intermittent quasi-laminar state even at $Re_\tau=60$ ($Re_m=1,720$).
Their computational box was $5\pi \delta \times 2\delta \times 2.5\pi \delta$.
Since it was smaller than that of the present one, a smaller box size might tend to retain the turbulent state down to a lower Reynolds number than a larger box.
Although the lower limit for sustaining turbulence cannot be clearly determined by the calculations up to now, the critical Reynolds number is estimated to lay between $Re_\tau=60$ and 64, namely $Re_{\rm m}=1,580$ and 1,860 or $Re_{\rm c}=930$ and 1,200 under the present condition. 

Transition experiments show that the plane channel flows typically undergo transition to turbulence at Reynolds numbers as low as $Re_{\rm c}=1,000$ \cite{Patel69,Carlson82}.
These experimental and the present numerical critical Reynolds numbers are both far below the critical Reynolds number of $Re_{\rm c}=5,780$ \cite{Thomas53,Orszag71} predicted with the two-dimensional linear stability analysis theory.
The discrepancy between the present critical Reynolds number ($\sim1,200$) and the experimental one ($\sim1,000$) may be attributed to an influence of periodic boundary condition in DNS and/or to that of sidewalls unavoidable in experiments. 

%%-------------------------------------- U_RMS
\begin{figure}[t]
 \begin{tabular}{cc}
  \hspace{-0.8em}
  \begin{minipage}[t]{0.45\textwidth}
   \begin{center}
    \epsfysize=57.5mm
    \epsfbox{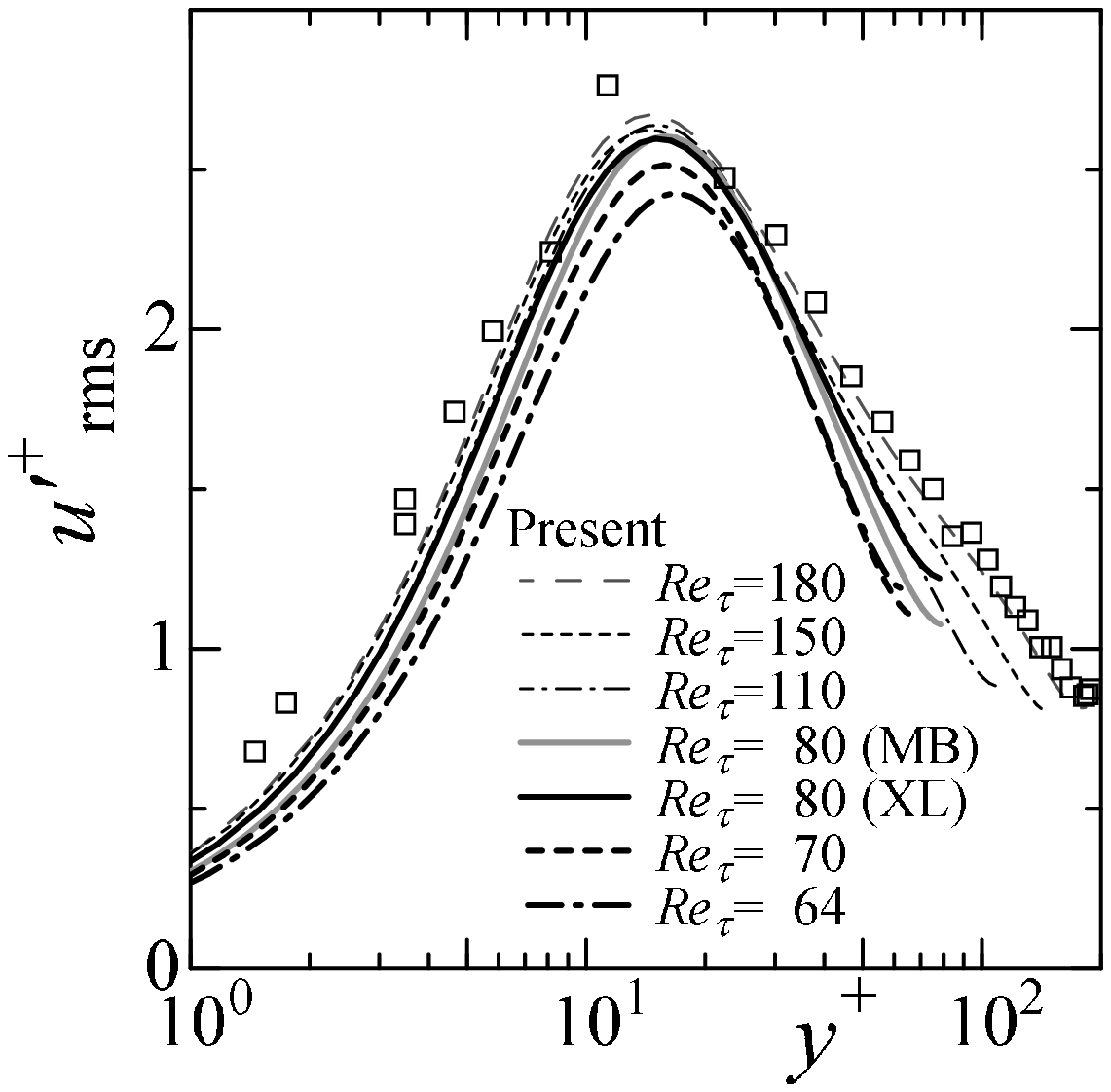}
   \end{center}
  \end{minipage}
  \hspace{-1.6em}
  \begin{minipage}[t]{0.60\textwidth}
   \begin{center}
    \epsfysize=57.5mm
    \epsfbox{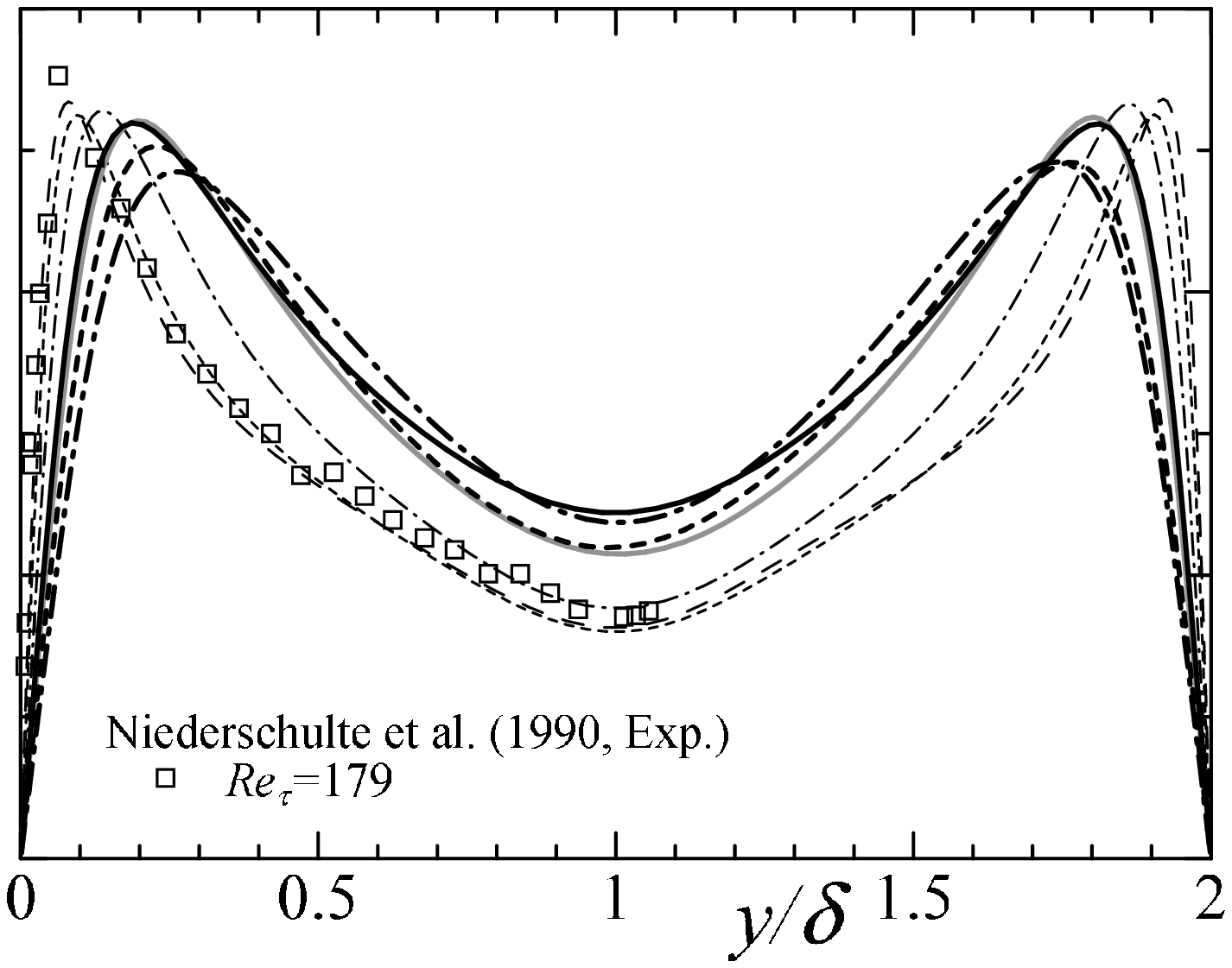}\\
   \end{center}
  \end{minipage}
 \end{tabular}\\
 \begin{center}
 \vspace{-1.0em}
 \caption{Root-mean-square of the streamwise velocity fluctuation $u'$.}
 \label{fig:urms}
 \end{center}
 \vspace{+0.7em}
%%\end{figure}

%%-------------------------------------- V_RMS, W_RMS
%%\begin{figure}[t]
 \begin{tabular}{cc}
  \hspace{-0.2em}
  \begin{minipage}[t]{0.49\textwidth}
   \begin{center}
    \epsfysize=33mm
    \epsfbox{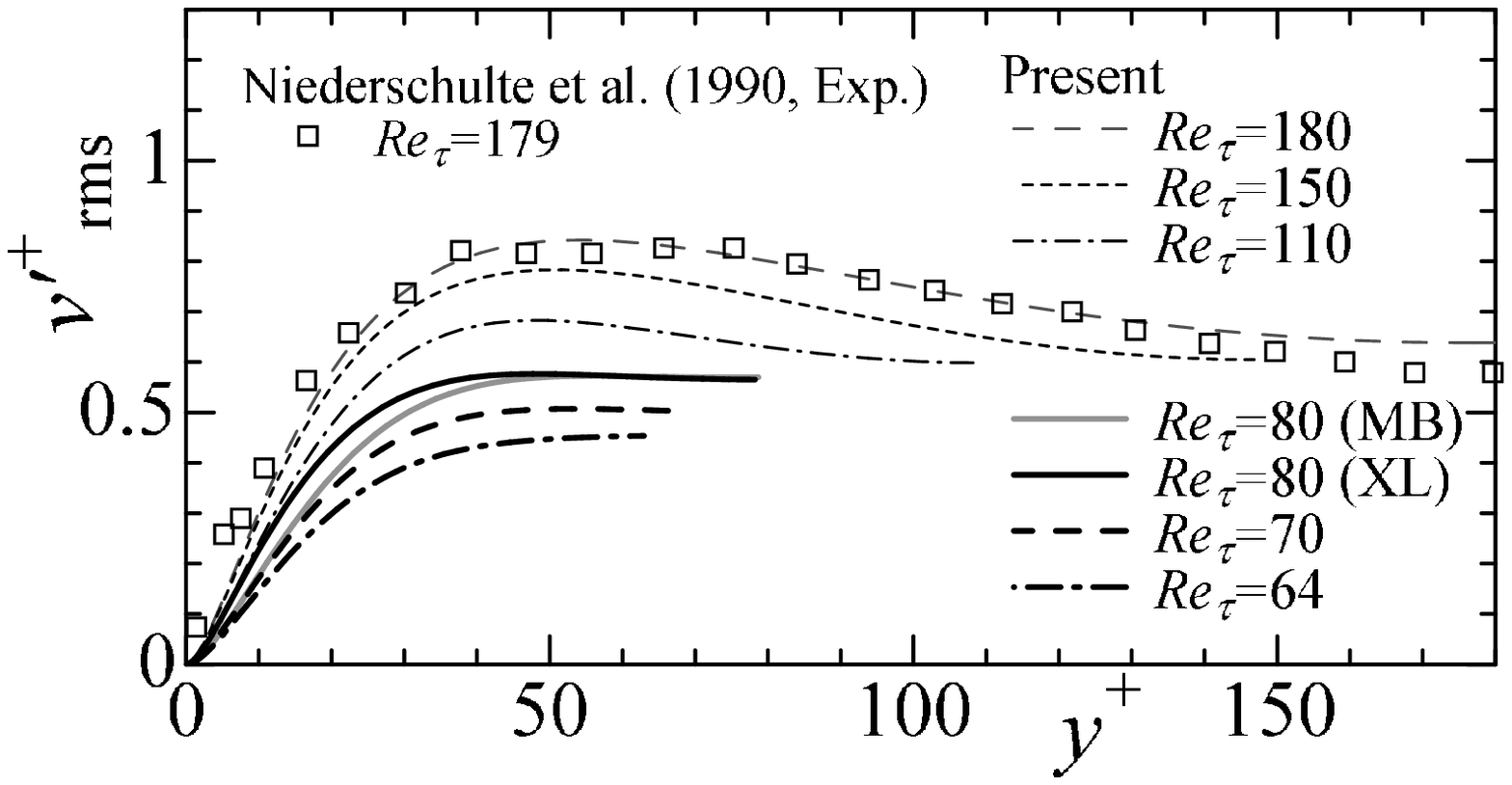}
   \end{center}
  \end{minipage}
  \hspace{-0.7em}
  \begin{minipage}[t]{0.49\textwidth}
   \begin{center}
    \epsfysize=33mm
    \epsfbox{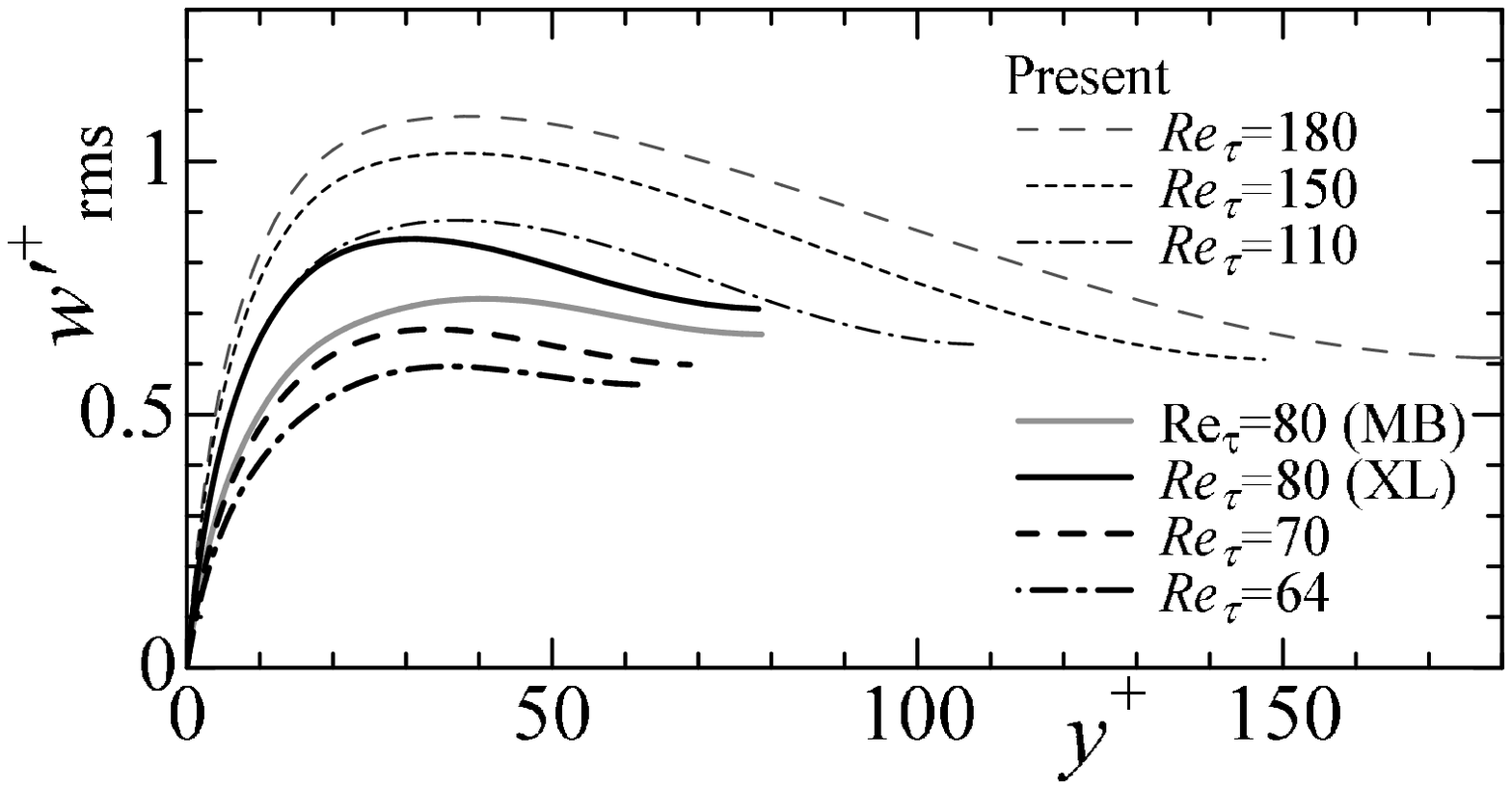}
   \end{center}
  \end{minipage}
 \end{tabular}\\
 \begin{center}
 \vspace{-4.5em}
 \hspace{+0.0em}(a) \hspace{14.6em} (b)\\
 \vspace{+2.0em}
 \caption{Root-mean-square of (a) the wall-normal fluctuation $v'$ 
          and (b) the spanwise fluctuation $w'$.
          }
 \label{fig:vw_rms}
 \end{center}
 \vspace{-0.2em}
\end{figure}

\subsection{Turbulence intensity}

The root-mean-square (r.m.s.) of the streamwise velocity fluctuation normalized by $u_\tau$ is given in \fref{fig:urms}. 
It increases remarkably at the centreline of the channel with decreasing the Reynolds number in the region of $Re_\tau<80$. 
The value of the local maximum near the wall decreases remarkably for $Re_\tau<80$.

With respect to the computational box size, a significant influence of the box size on $u'^+_{\textrm{\tiny{rms}}}$ can be seen for $Re_\tau=80$ at the central region. 
With extending box size, $u'^+_{\textrm{\tiny{rms}}}$ is enhanced, with a deviation of about 12\% 
at the channel centre. 
This tendency is also seen for the spanwise fluctuations in the whole channel (\fref{fig:vw_rms}(b)), with a deviation of 7--27\%. 
The influence on the wall-normal component is small (\fref{fig:vw_rms}(a)).
Moreover, the influence of the box size on the mean flow variables and other turbulence quantities, such as $\overline{u}^+$ and vorticity fluctuations (shown later), are also significant at the very low Reynolds number of $Re_\tau=80$, whereas the those of the box size on the turbulence statistics are rather small for the moderate Reynolds number of $Re_\tau=180$--640 as shown by Abe \etal \cite{Abe04a}.

Effects of Reynolds number significant in the ${u_i}'^+_{\textrm{\tiny{rms}}}$ of both the spanwise and the wall-normal directions.
All the component values decrease as the Reynolds number decreases. 
Antonia \etal \cite{Antonia92} indicated that the Reynolds-number dependence of $w'^+_{\textrm{\tiny{rms}}}$ is more significant compared to these of $u'^+_{\textrm{\tiny{rms}}}$ and $v'^+_{\textrm{\tiny{rms}}}$. 
In the present work, both $v'^+_{\textrm{\tiny{rms}}}$ and $w'^+_{\textrm{\tiny{rms}}}$ decrease remarkably with $Re_\tau$. 
This is because the production term of $u'^+_{\textrm{\tiny{rms}}}$ and the redistributions for $v'^+_{\textrm{\tiny{rms}}}$ and $w'^+_{\textrm{\tiny{rms}}}$ are reduced with the decreasing $Re_\tau$ as 
discussed later. 

%%====================================== TABLE: NEAR-WALL
\begin{table}[t]
  \tbl{Near-wall expansion coefficient}
 {\begin{tabular}{cl ccc ccc}\toprule
$Re_\tau$     & \hspace{-1.2em}   & $b_1$ & ~$c_2$                
                                  & $b_3$ & ~$\overline{b_{1,3}-b_{3,1}}$ 
                                          & ~$\overline{b_1c_2}$ 
                                          & $(\overline{b_1d_2}+\overline{c_1c_2})$ \\\colrule
180           & \hspace{-1.2em}   & 0.360 & ~$8.70\times10^{-3}$ 
                                  & 0.189 & ~$2.65\times10^{-2}$ 
                                          & ~$7.29\times10^{-4}$ & ~$2.0\times10^{-5}$ \\
~180$^{\rm a}$& \hspace{-1.2em}   & 0.356 & ~$8.5~\,\times10^{-3}$
                                  & 0.190 & ~$2.7~\,\times10^{-2}$ 
                                          & ~$7.0~\,\times10^{-4}$ & ----- \\
150           & \hspace{-1.2em}   & 0.354 & ~$8.60\times10^{-3}$ 
                                  & 0.172 & ~$2.60\times10^{-2}$ 
                                          & ~$7.08\times10^{-4}$ & ~$1.9\times10^{-5}$ \\
110           & \hspace{-1.2em}   & 0.336 & ~$6.75\times10^{-3}$ 
                                  & 0.145 & ~$2.46\times10^{-2}$ 
                                          & ~$5.41\times10^{-4}$ & ~$1.5\times10^{-5}$ \\
~\,80         & \hspace{-1.2em}MB & 0.302 & ~$4.55\times10^{-3}$ 
                                  & 0.105 & ~$2.06\times10^{-2}$ 
                                          & ~$3.42\times10^{-4}$ & ~$0.8\times10^{-5}$ \\
~\,80         & \hspace{-1.2em}XL & 0.333 & ~$6.85\times10^{-3}$ 
                                  & 0.144 & ~$2.28\times10^{-2}$ 
                                          & ~$4.45\times10^{-4}$ & ~$2.0\times10^{-5}$ \\
~\,70         & \hspace{-1.2em}   & 0.291 & ~$4.65\times10^{-3}$ 
                                  & 0.101 & ~$2.08\times10^{-2}$ 
                                          & ~$3.11\times10^{-4}$ & ~$0.9\times10^{-5}$ \\
~\,64         & \hspace{-1.2em}   & 0.268 & ~$3.90\times10^{-3}$ 
                                  & 0.086 & ~$1.91\times10^{-2}$ 
                                          & ~$2.45\times10^{-4}$ & ~$0.6\times10^{-5}$ \\\botrule
  \end{tabular}}
 \tabnote{$^{\rm a}$Antonia \& Kim \cite{Antonia94}}
 \label{tab:nw}
\end{table}
%%=======================================================

\begin{figure}[tbp]
 \begin{center}
  \hspace{-1.5em}
    \epsfysize=55.5mm
    \epsfbox{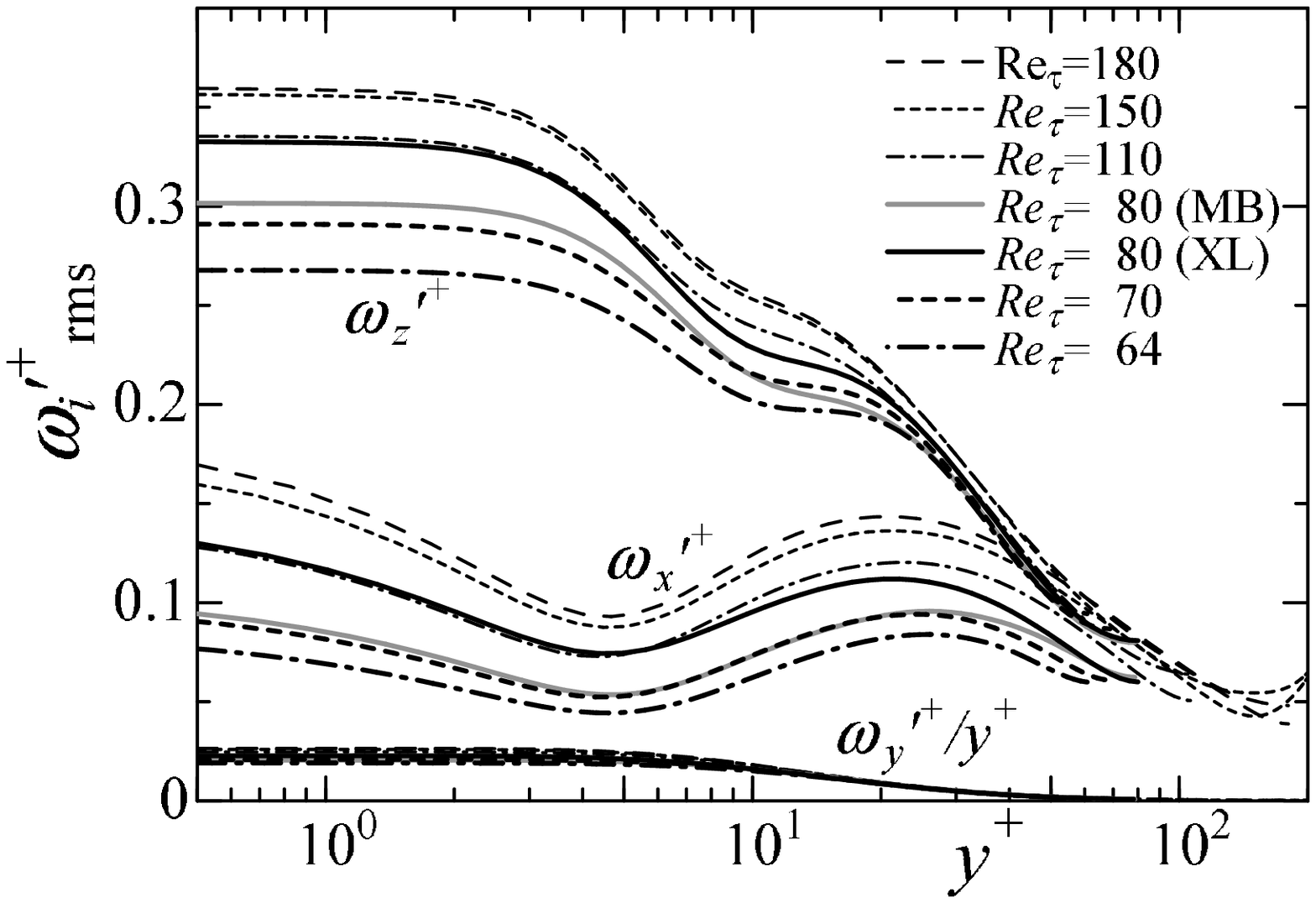}
    \epsfysize=55.5mm
    \epsfbox{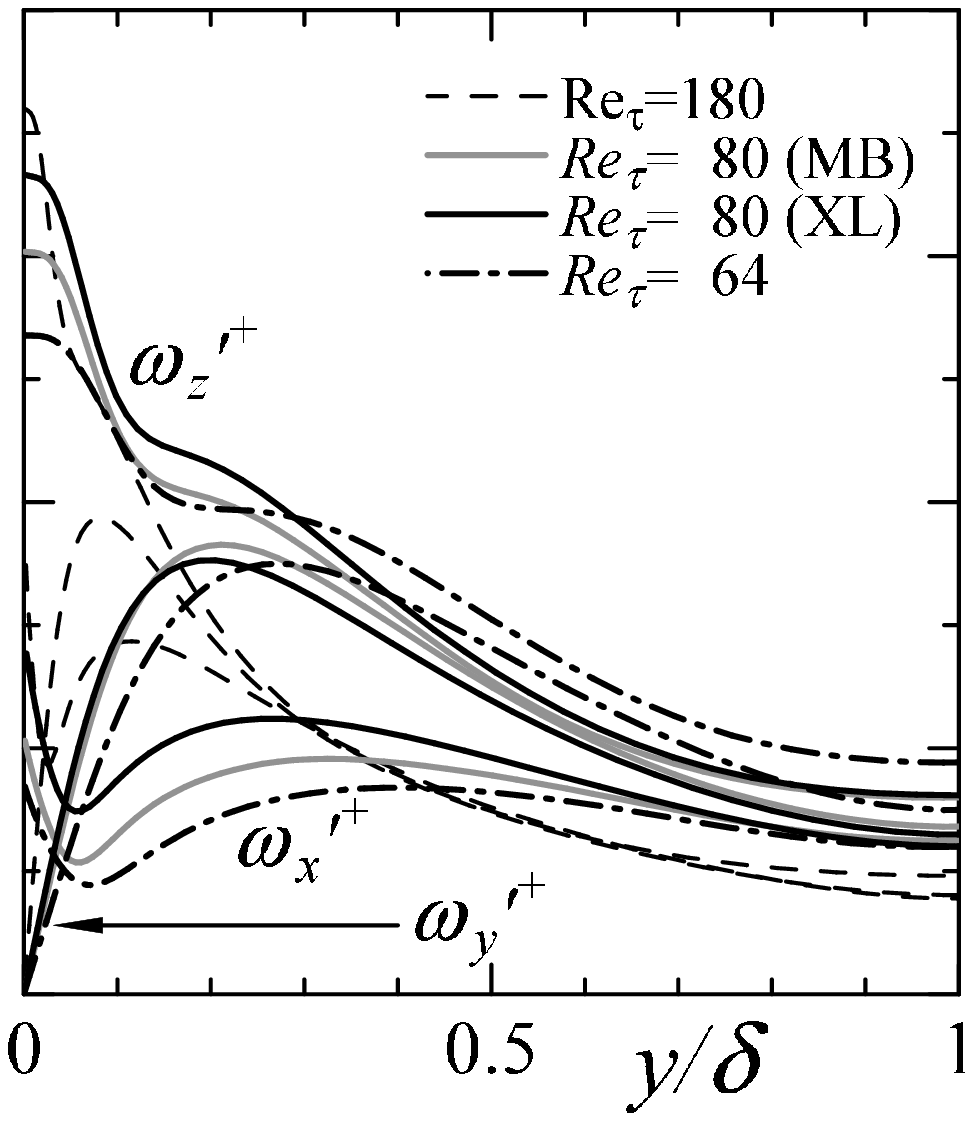}\\
 \vspace{-0.5em}
 \caption{Root-mean-square of vorticity fluctuation.}
 \label{fig:omg}
 \end{center}
 \vspace{-0.5em}
\end{figure}

To analyse the near-wall asymptotic behaviour, the velocity fluctuations can be expanded in Taylor series about-the-wall value as follows: 
\begin{equation}
 \left\{
  \begin{array}{@{\,}ll}
   u'^+=\hspace{0.2em}b_1 y^+ +c_1 y^{+2}+d_1 y^{+3}+\cdots \\
   v'^+=\hspace{3.5em}         c_2 y^{+2}+d_2 y^{+3}+\cdots \\
   w'^+=              b_3 y^+ +c_3 y^{+2}+d_3 y^{+3}+\cdots. \\
  \end{array}
 \right.
\label{eq:nw}
\end{equation}
The values of the coefficients of the first terms in \eref{eq:nw} are shown in \tref{tab:nw}.
The r.m.s. values of the vorticity fluctuations are shown in \fref{fig:omg}. 
The streamwise and spanwise vorticity fluctuations, namely $\omega'^+_x(=b_3$) and $\omega'^+_z(=b_1)$, decrease with decreasing Reynolds number. 
However, the influence of insufficient box size for $Re_\tau=110$ and 80~(MB) cannot be neglected. 
The ratio $\omega'^+_y/y^+(=b_{1,3}-b_{3,1} \equiv \partial b_1/\partial z^+ + \partial b_3/\partial x^+$) tends to become constant in the near-wall region as reported by Antonia \& Kim \cite{Antonia94}. 

%%-------------------------------------- R.M.S. OF VORTICITY

The use of \eref{eq:nw} in the expression for the Reynolds shear stress, $-\overline{u'^+v'^+}$, yields 
\begin{equation}
 -\overline{u'^+v'^+}=-\Big[ \overline{b_1c_2}y^{+3}
                      +\big(\overline{b_1d_2}+\overline{c_1c_2} \big)y^{+4}
                      +\cdots \Big].
\label{eq:exp2}
\end{equation}
The near-wall values of $(\overline{b_1c_2})$ and $(\overline{b_1d_2}+\overline{c_1c_2})$ are extrapolated up to the wall, and are also given in \tref{tab:nw}. 

The present results at $Re_\tau=180$ agree well with those of Antonia \& Kim \cite{Antonia94}. 
All of the coefficients, shown in \tref{tab:nw}, decrease with the decrease of the Reynolds number. 
This is because the production rate of the turbulent kinetic energy decreases with decreasing the Reynolds number as discussed in \sref{shear}. 
The decrease in $b_1$ with $Re_\tau$ is smaller than that in either $c_2$ or $b_3$, which is comparable to the tendency toward more anisotropic turbulence at lower Reynolds numbers, cf. figures \ref{fig:urms} and \ref{fig:vw_rms}.
When the Reynolds number is decreased from 180 to 64, 
the decrease in $b_1$ is only 25\%, 
compared with 55\% in the cases of $c_2$ and $b_3$.
The decreases of $c_2$, $b_3$, $\overline{b_1c_2}$ and $(\overline{b_1d_2}+\overline{c_1c_2})$ are significant when the Reynolds number falls to 70 or 64. 
The dependence of the ratio $\omega'^+_y/y^+(=\overline{b_{1,3}-b_{3,1}})$ on the Reynolds number is not large but still appreciable, i.e., drop of 28\% with $Re_\tau =180\to 64$.

In the case of $Re_\tau\leq 80$~(MB), however, the obtained coefficients could be further influenced by the box size. 
For all the values, the increases due to extending the box size are not negligible compared to the variation with the Reynolds number.

\subsection{Reynolds shear stress}
\label{shear}

%%-------------------------------------- REYNOLDS SHEAR STRESS
\begin{figure}[t]
 \centerline{\epsfxsize=120.0mm
    \epsfbox{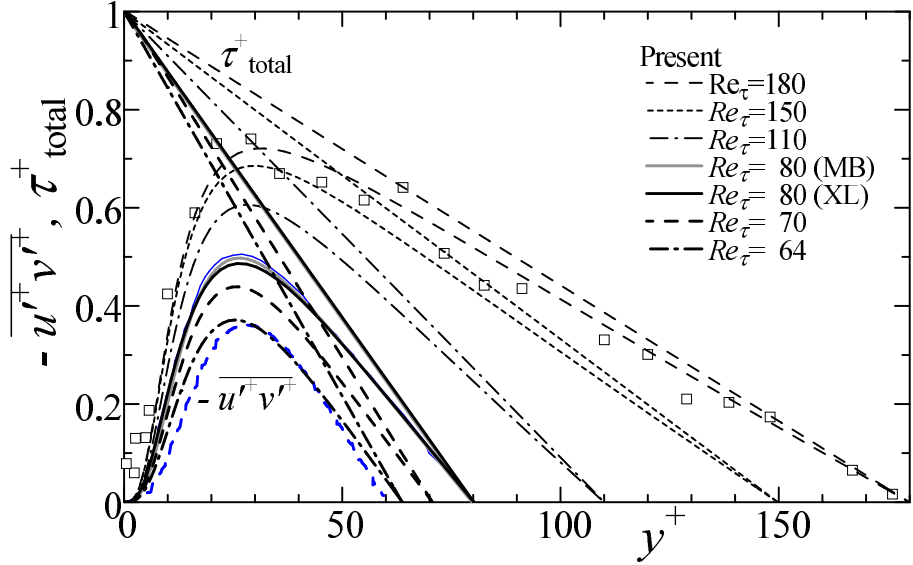}}
    \vspace{-0.0em}
    \caption{Reynolds shear stress $-\overline{u'^+v'^+}$ and total shear 
             $\tau_{\rm total}=-\overline{u'^+v'^+}+{\rm d}\overline{u}^+/{\rm d}y^+$ distribution.
             Blue lines, DNS by Iida \& Nagano \cite{Iida98} at
             $Re_\tau=80$ (\full) and $Re_\tau=60$ (\broken);
             symbol (\opensquare), experiment by Niederschulte \etal \cite{Niederschulte90}
             at $Re_\tau=179$.
             }
  \label{fig:shear}
  \vspace{+0.5em}
\end{figure}

\Fref{fig:shear} shows the Reynolds shear stress $-\overline{u'^+v'^+}$ and the total shear stress $\tau_{\tiny{\textrm{total}}}$. 
In all the cases of the present calculations, the profile of the total shear stress is given as a straight line, indicating that the flow reaches a statistically steady state. 
As the Reynolds number decreases, the peak value of $-\overline{u'^+v'^+}$ decreases and its position moves close to the wall, if scaled with the wall unit. 
When $Re_\tau$ is 180, the peak of $-\overline{u'^+v'^+}$ reaches 0.72 at $y^+=32$, while it becomes 0.37 at $y^+=26$ in the case of $Re_\tau=64$. 

%%-------------------------------------- PROD. OF K
\begin{figure}[t]
 \begin{center}
  \epsfxsize=120.0mm
  \epsfbox{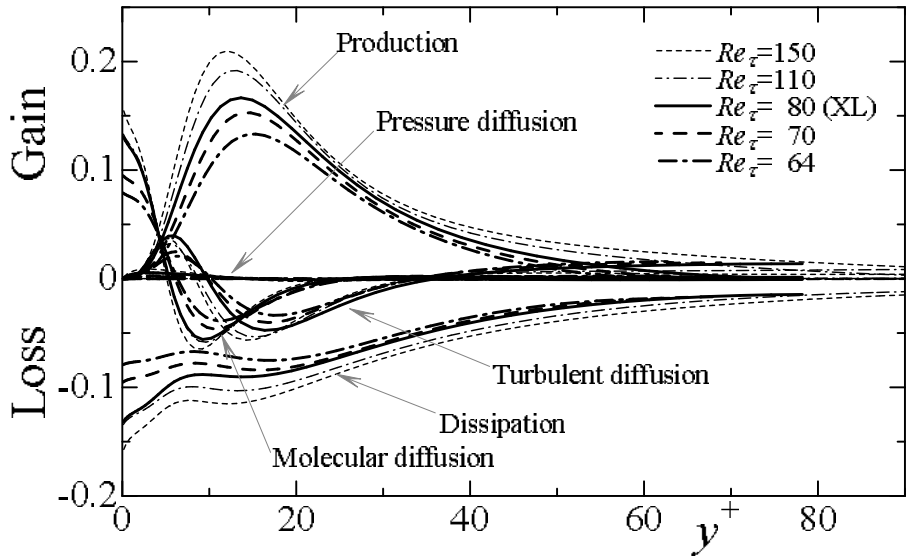}
  \vspace{-0.3em}
 \caption{Budget of turbulent kinetic energy $k$ normalized by $\nu/u_\tau^4$}.
 \label{fig:budget}
 \end{center}
\end{figure}

%%-------------------------------------- PEAK OF PROD. OF K
\begin{figure}[h]
 \begin{center}
  \epsfxsize=120.0mm
  \epsfbox{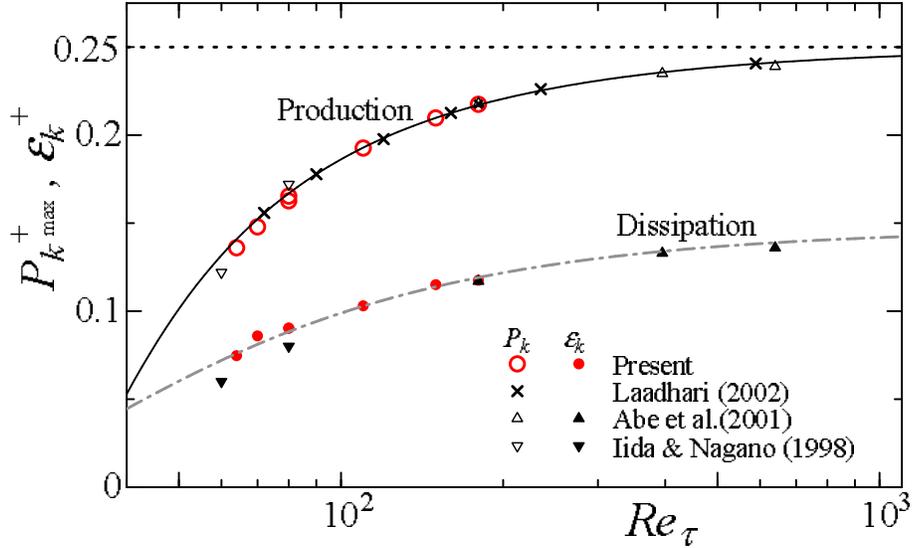}\\
 \caption{Variation of the peak of production term $P_{k,\textrm{\tiny{max}}}$ 
          and the dissipation rate $\varepsilon_k$ at the same height.
          The fitting curve of ({\protect\mbox{------}}) 
          is the empirical correlation, cf. \cite{Laadhari02}.
          }
 \label{fig:budget_peak}
 \end{center}
\end{figure}

In the fully-developed channel flow, the production term of the turbulent kinetic energy is expressed as 
\begin{equation}
 P_k^+=-\overline{u'^+v'^+} \frac{\partial \overline{u}^+}{\partial {y^+}}
 =-\overline{u'^+v'^+}\bigg[1-\frac{y^+}{Re_\tau}-\big(-\overline{u'^+v'^+}\big)\bigg].
\label{eq:prod}
\end{equation}
Thus the peak value of $P_k$ can be calculated as 
\begin{equation}
 P_k^+=\frac{1}{4}\bigg[1-\frac{y^+_{\tiny{\textrm{max}}}}{Re_{{{\tau}}}}\bigg]^2
 \hspace{0.5em} \textrm{at} \hspace{0.2em} 
 \bigg\{y^+\bigg|
 -\overline{u'^+v'^+} = \frac{\partial \overline{u}^+}{\partial {y^+}} 
 \bigg\}.
\label{eq:prod_max}
\end{equation}
Here, the wall-normal position of the peak of $P_k^+$ is denoted as $y^+_{\tiny{\textrm{max}}}$, at which the Reynolds shear stress and the viscous stress are equal to half of the total shear stress.

In the present calculation, $y^+_\textrm{\tiny{max}}$ is 12.5 for $Re_\tau=180$. 
When the Reynolds number is decreased, it moves away from the wall and reaches at $y^+_\textrm{\tiny{max}}\approx19.5$ for $Re_\tau =64$. 
Sahay \& Sreenivasan \cite{Sahay99} and Laadhari \cite{Laadhari02} examined the Reynolds-number dependence of $y^+_{\tiny{\textrm{max}}}$, and related it to those positions at which turbulence production and momentum transport attained their respective maxima. 
They investigated the evolution of $y^+_{\tiny{\textrm{max}}}$ in the form $y^+_{\tiny{\textrm{max}}}=y^+_{\tiny{\textrm{max}}_\infty} + f(Re_\tau)$ where $y^+_{\tiny{\textrm{max}}}\to y^+_{\tiny{\textrm{max}}_\infty}$ ($Re_\tau \to \infty$) is an average value of $y^+_{\tiny{\textrm{max}}_\infty}=11$ from DNS data for channel flows. 
A power law fitting of the present data gives 
\begin{equation}
  y^+_{\tiny{\textrm{max}}} \approx 11+ \bigg(\frac{243}{Re_\tau}\bigg)^\frac{3}{2}.
\label{eq:y_max}
\end{equation}

The peak value of $P_k^+$ decreases with decreasing $Re_\tau$ (see \fref{fig:budget}). 
\Fref{fig:budget_peak} shows the peak value of $P_k^+$ and dissipation rate $\varepsilon_k^+$ at $y^+_\textrm{\tiny{max}}$ for each Reynolds number. 
The empirical relation between $P_{k,\textrm{\tiny{max}}}^+$ and $Re_\tau$, derived by Laadhari \cite{Laadhari02}, 
shows a good agreement with the present DNS results.
The decrease in $P^+_{k,\textrm{\tiny{max}}}$ with decreasing $Re_\tau$ is clearly more prominent than that of $\varepsilon_k^+$. 
This is a reason why the turbulent intensities are decreased in the near-wall region with the decreasing $Re_\tau$ as seen in figures \ref{fig:urms}, \ref{fig:vw_rms} and in \tref{tab:nw}. 

\subsection{Turbulence structures: coherent structure and LSM}
\label{sec:structure}

\subsubsection{Two-point correlation coefficient}

\begin{figure}[t]
 \begin{tabular}{cc}
 \hspace{-0.8em}
 \begin{minipage}[t]{0.49\textwidth}
  \begin{center}
   \epsfysize=46.0mm
   \epsfbox{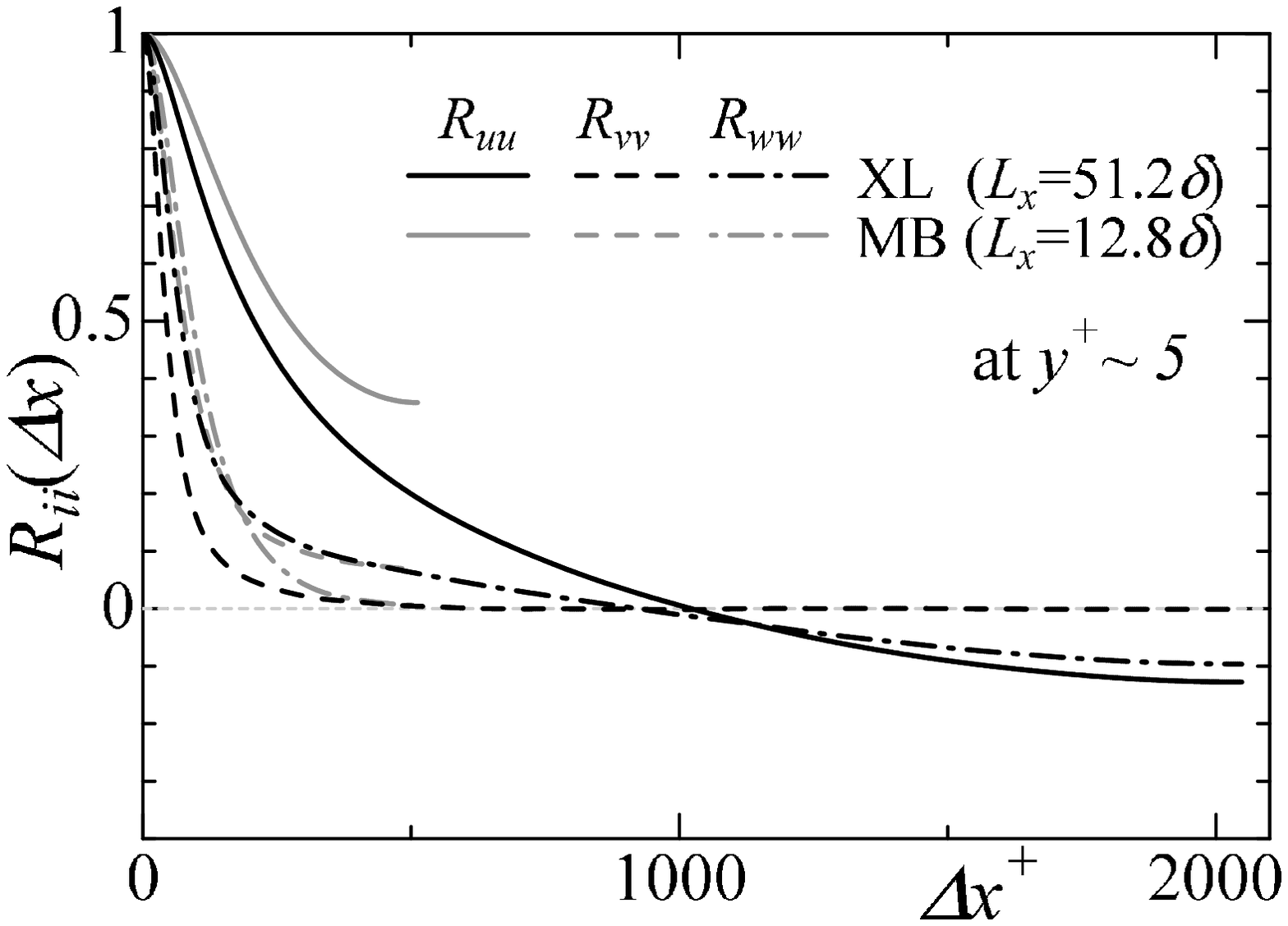}
  \end{center}
 \end{minipage}
 \hspace{-0.4em}
 \begin{minipage}[t]{0.49\textwidth}
  \begin{center}
   \epsfysize=46.0mm
   \epsfbox{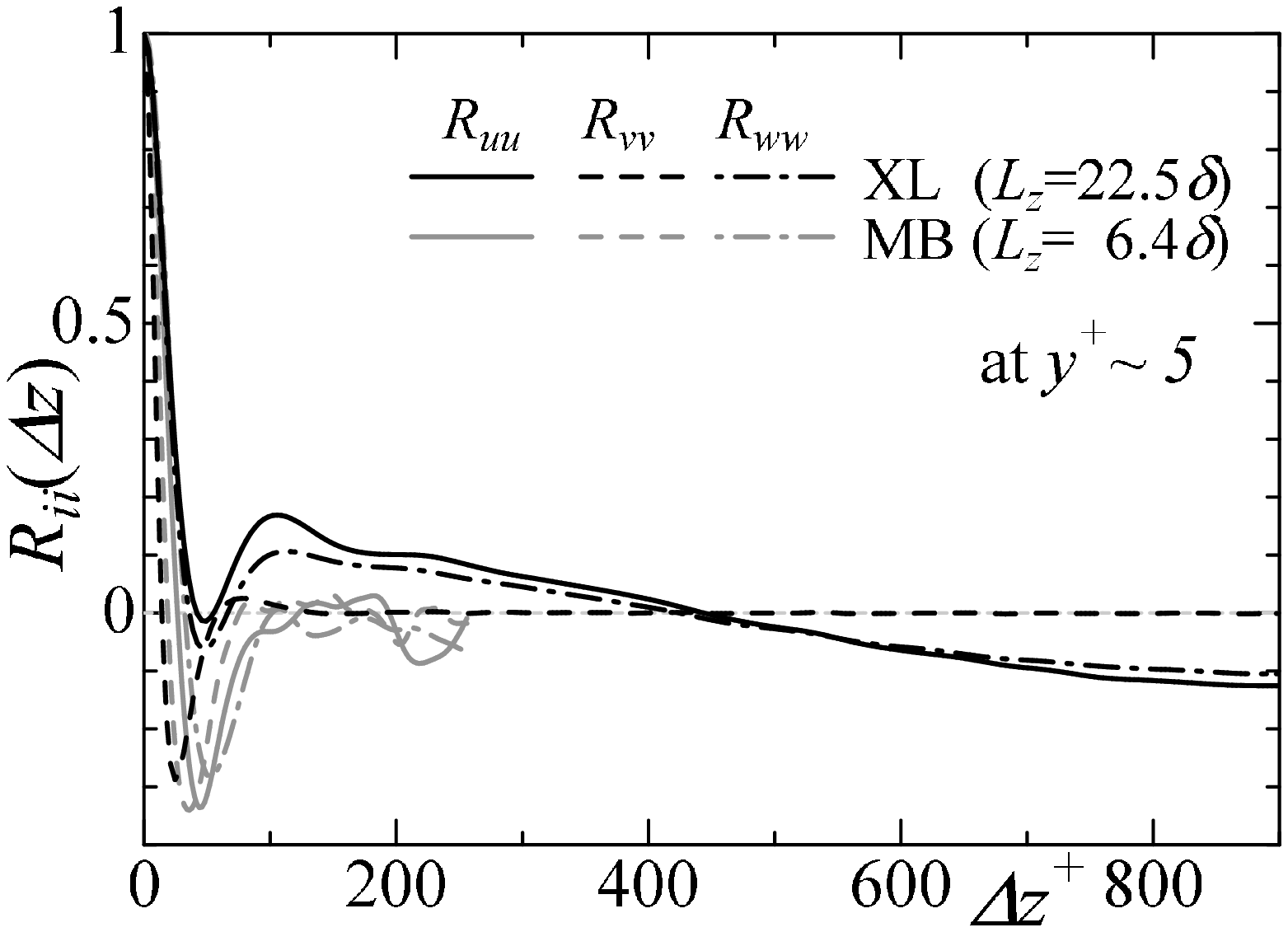}
  \end{center}
 \end{minipage}
 \end{tabular}\\
 \begin{center}
 \vspace{-7.2em}
 \hspace{+10.0em}(a) \hspace{14.8em} (b)\\
 \vspace{+5.5em}
 \caption{Two-point correlation coefficients $R_{ii}$ of the velocity fluctuation component at $Re_\tau=80$ 
          with medium or extra-large box size;
          (a) streamwise, (b) spanwise.
          }
 \label{fig:tpc}
 \end{center}
\end{figure}

The effect of the box size can be most clearly observed in the streamwise and spanwise two-point correlations of the velocity fluctuations $R_{ii}$, which is shown in \fref{fig:tpc}. 
The data of $R_{ii}$ were obtained in a near-wall region at $y^+\approx 5$ for $Re_\tau=80$. 
In the case of the medium box (MB), $R_{uu}$ does not fall down to zero, indicating that the box size is not large enough: especially, the streamwise box length is too short to contain the streaky structures in the near-wall region (see \fref{fig:tpc}(a)).
On the other hand, $R_{uu}$ and $R_{ww}$ fall down to negative values at the maximum separation (half the domain size) for both stream- and the spanwise directions for the large box (XL), while $R_{vv}$ falls off to almost zero. 
It indicates that one long wavelength structure is captured with XL. 
In addition, a significant decrease in the magnitude of the negative maximum is found at the spanwise separation distance of $\Delta z^+ \approx 50$, which corresponds to the spanwise spacing of the near-wall streaky structures (\fref{fig:tpc}(b)). 
The negative values of $R_{uu}$ and $R_{ww}$ at the mid box length become the largest in the channel centre region (not shown here).
These suggest that an influence of a structure that is larger than a streaky structure exists not only in the channel centre but also in the near-wall region.
The spacing of the streaky structures and the large-scale structures are investigated with the use of the pre-multiplied energy spectra, which will be discussed in more detail.

%%-------------------------------------- T.P.C.
\subsubsection{Pre-multiplied energy spectra}
\label{sec:pmes}

\begin{figure}[t]
 \begin{tabular}{cc}
  \hspace{-0.2em}
  \begin{minipage}[t]{0.49\textwidth}
   \begin{center}
    \epsfysize=55.0mm
    \epsfbox{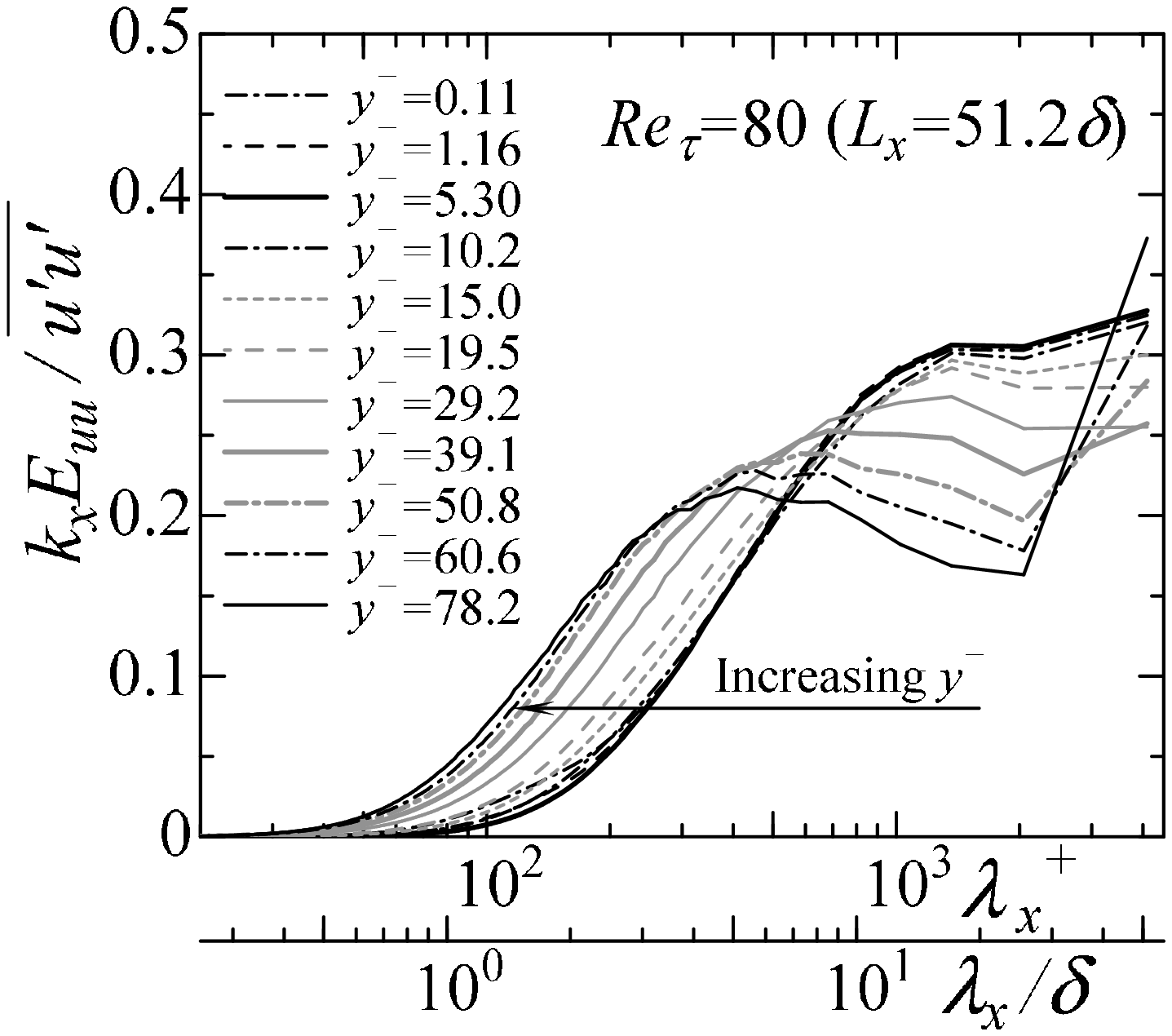}
   \end{center}
  \end{minipage}
  \hspace{-0.5em}
  \begin{minipage}[t]{0.49\textwidth}
   \begin{center}
    \epsfysize=55.0mm
    \epsfbox{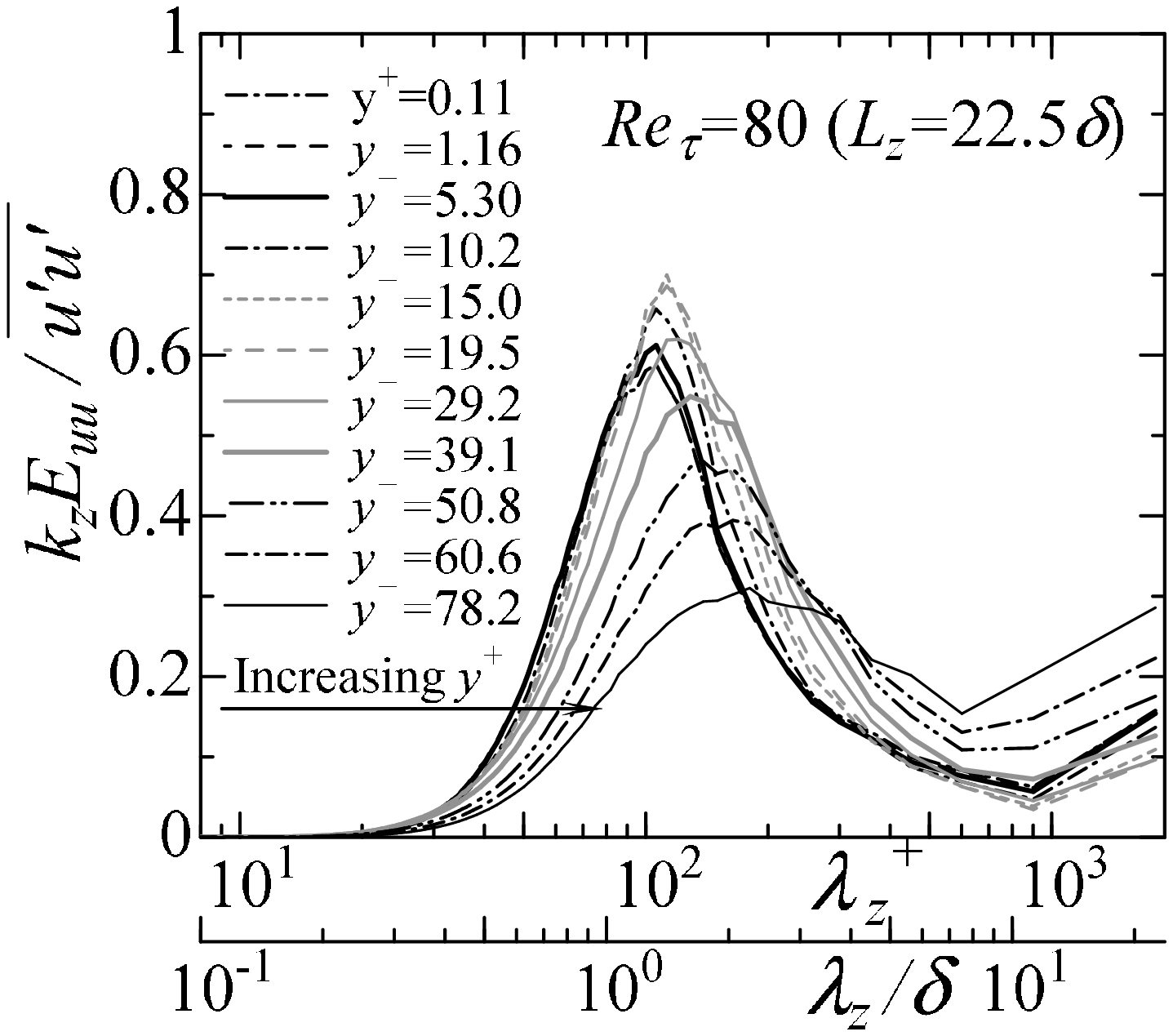}
   \end{center}
  \end{minipage}
 \end{tabular}
 \begin{center}
 \vspace{-12.2em}
 \hspace{+8.0em}(a) \hspace{14.8em} (b)\\
 \vspace{+10.5em}
 \caption{Pre-multiplied energy spectra for $Re_\tau=80$ 
          with extra-large box size of ($L_x \times L_z$) = ($51.2\delta \times 22.5\delta$);
          (a) streamwise $k_xE_{uu}/\overline{u'u'}$, (b) spanwise $k_zE_{uu}/\overline{u'u'}$.
          }
 \label{fig:pmes}
 \end{center}
\end{figure}

The streamwise and spanwise pre-multiplied energy spectra for $Re_\tau =80$~(XL) are shown in \fref{fig:pmes} with reference to Jim\'enez \cite{Jimenez98}. 
To examine the hypothesis that the higher-Reynolds-number turbulence structure differs significantly from that at low Reynolds numbers, spectra of $u$ and $v$ were examined by the experiment of Wei \& Willmarth \cite{Wei89} and DNS of other researchers \cite{Abe04a,Jimenez98,Antonia92}.
Their spectra were presented at several wall-normal positions, generally in the form of $k_i^+ E_{uu}$ \emph{vs.} $\log \lambda_i^+$, where $k_i$ and $\lambda_i \equiv 2\pi/k_i$ are  wavenumber and wavelength, respectively, and the normalization for the wavenumber power spectral $E_{uu}$ is such that 
\begin{equation}
  \int_0^\infty  E_{uu}(k_x){\rm d}k_x
 =\int_0^\infty  E_{uu}(k_z){\rm d}k_z = \overline{u'u'}.
\end{equation}
When the spectrum is plotted in the form $k_i E_{uu}$, the area under the curve are proportional to the contribution to the total energy from the wavenumber bands. 
Since the pre-multiplied spectrum is proportional to the power in a logarithmic band at $k_i$ or $\lambda_i$ (see, e.g., Perry \etal \cite{Perry86}),
the peak position of the pre-multiplied energy spectra gives the most energetic wavelength (MEW). 

It is interesting to note that MEW in the wall vicinity stays still at about $\lambda_z^+\approx 100$ and not much changed from that of the higher Reynolds numbers even though $Re_\tau$ is as low as 80 (\fref{fig:pmes}(b)). 
This corresponds to the well-known streaky structure in a wall vicinity of wall-bounded turbulence.
Moving away from the wall, the MEW shifts slightly to the longer wavelengths: $\lambda_z^+\approx 130$ or $\lambda_z\approx 1.6\delta$ at the mid height.
In the central region, a peak of MEW arises at about $\lambda_z^+=180$. 
This MEW corresponds to about 2.3$\delta$, which shows a significant deviation from 1.3--1.6$\delta$ obtained for the higher Reynolds numbers, cf. Abe \etal \cite{Abe04a}. 
The streamwise MEW arises at $\lambda_x^+\approx1,000$ in the near-wall region. 
With increase in the distance from the wall, the streamwise MEW moves towards the shorter wavelength of $\lambda_x^+\approx400$ or $\lambda_x \approx 0.5\delta$ (\fref{fig:pmes}(a)). 

%%-------------------------------------- PRE-MULTIPLIED SPECTRA

On the other hand, in the both directions, another peak appears at the longest wavelength in the core region, indicating consistency with the two-point velocity correlation $R_{uu}$, shown in \fref{fig:tpc}.
The substantial secondary peaks of $w$ (the figure is not shown here) with wavelengths of $\lambda_x=L_x$ and $\lambda_z=L_z$ are also observed, this is an evidence that there exist a large-scale motion in spanwise direction.

\subsubsection{Instantaneous flow field}

%%-------------------------------------- INSTANTANEOUS FLOW FIELD
\begin{figure}[tbp]
 \centerline{\epsfxsize=128.5mm
  \epsfbox{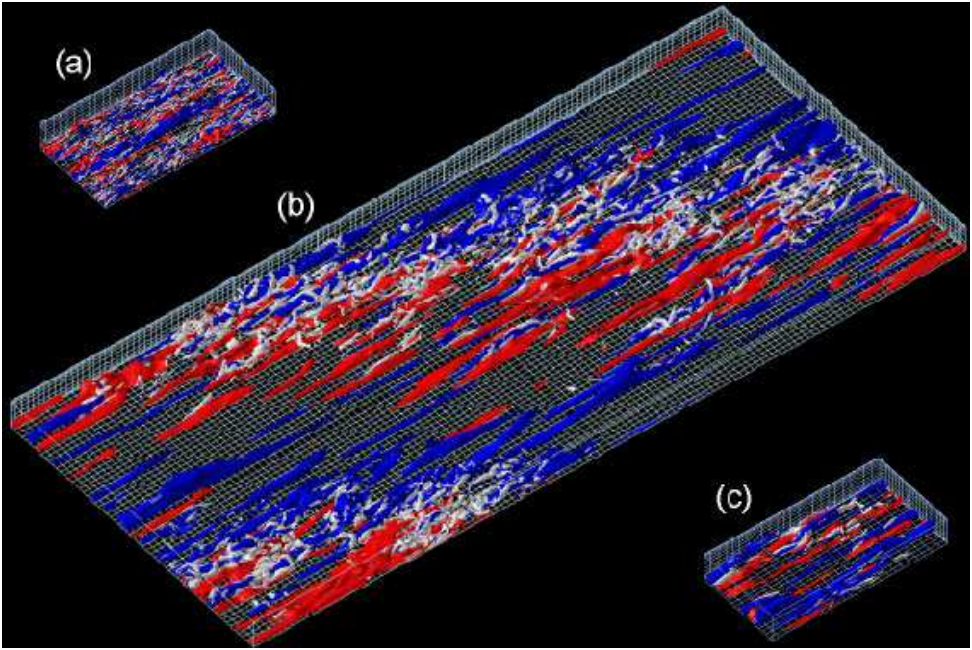} \\}
 \vspace{0.3em}
 \caption{Instantaneous flow fields; high- (red) and low-speed (blue) regions of $u'^+$ 
          and negative regions (white) of the second invariant of deformation tensor 
          $u'_{i,j}u'_{j,i}$ which is equivalent to the vortical position.
          (a) $Re_\tau =180$, (b) $Re_\tau =80$~(XL), (c) $Re_\tau =80$~(MB).
          All of the box size (a)--(c) are scaled by $\delta$.
          Direction of the mean flow is from bottom-left to top-right. 
          The visualized volume is the lower half of the computational box, 
          namely (a,c) $12.8\delta \times \delta \times 6.4\delta$ or 
          (b), $51.2\delta \times \delta \times 22.5\delta$.
          }
 \label{fig:instfld}
 \vspace{0.3em}
\end{figure}

\Fref{fig:instfld} shows the high- and low-speed regions and the second invariant of deformation tensor ($I\hspace{-0.2em}I'=\partial u'_i/\partial x_j \times \partial u'_j/\partial x_i$) at $Re_\tau=180$, 80~(MB) and 80~(XL). 
For $Re_\tau=180$ and 80~(MB), the high- and low-speed streaks are evenly distributed as seen in figures~\ref{fig:instfld}(a) and (c).
Note that the elongated streaks at $Re_\tau=80$ penetrate the computational domain of MB as shown in  \fref{fig:instfld}(c)), indicating a shortage of the box size.
The enlarged box-size XL is valid for capturing them, as seen in \fref{fig:instfld}(b).
For $Re_\tau=80$~(XL), on the contrary, the near-wall streaks are intermittently distributed, and a rather calm region can be observed. 
If we pay attention to the localized cluster of streaks, the spanwise spacing of low-speed streaks is essentially invariant with Reynolds number and the box size, exhibiting consistent value of $\Delta z^+=100$.
The flow visualization study of Carlson \etal \cite{Carlson82}, on the transition structure at $Re_c=1,200$, revealed that natural turbulence spots appeared randomly across the span and the additional oblique waves play an important role in the breakdown to turbulence.
Both of their Reynolds number and the shape of the oblique structure are consistent with those we have obtained.

The well-known vortex structures such as quasi-streamwise vortexes are dominant and exist in the buffer region. 
In addition, these vortices are associated closely with the crowded near-wall streaks. 
It may be observed that strong turbulence-production regions, ejections and sweeps, in the buffer layer appear in the crowded streaks and vortexes area.
The long-wavelength structure of the strong/weak-turbulent regions occurs periodically. 
Its streamwise and spanwise wavelengths occupy almost the whole box lengths on each direction. 
Moreover, the streamwise velocity profile changes from a more flat, turbulent-like profile, to a laminar Poiseuille-like profile inside the weak-turbulent region. 
In other words, the flow field can be separated into two areas: a upstream quasi-laminar state and a downstream turbulent state, which are relatively high- and low-speed regions, respectively.
Accordingly, the relatively high-speed region overtakes the low-speed one, which produces the strong turbulence and resultant puff-like structure at the interface, as discussed in the following section.
This is a reason why the streamwise two-point correlations $R_{uu}$ falls down to a negative value at the middle of the box as seen in \fref{fig:tpc}. 
It is interesting to note that, with using a computational domain as large as XL, it enables us to capture the large-scale oblique structure --- consisting of quasi-laminar and strong turbulent states --- which has been unable to emerge in a usual computational box size, i.e., MB and LB.
From DNS with a domain similar size to MB, neither the quasi-laminar nor the turbulent state are stable at a very low Reynolds number, and in a relatively short time the flow repeatedly returns to the other state (as discussed by Iida \& Nagano \cite{Iida98}).

%%-------------------------------------- FLATNESS FACTOR
\begin{figure}[tbp]
 \begin{tabular}{cc}
  \hspace{-0.5em}
  \begin{minipage}[t]{0.49\textwidth}
   \begin{center}
    \epsfysize=45.0mm
    \epsfbox{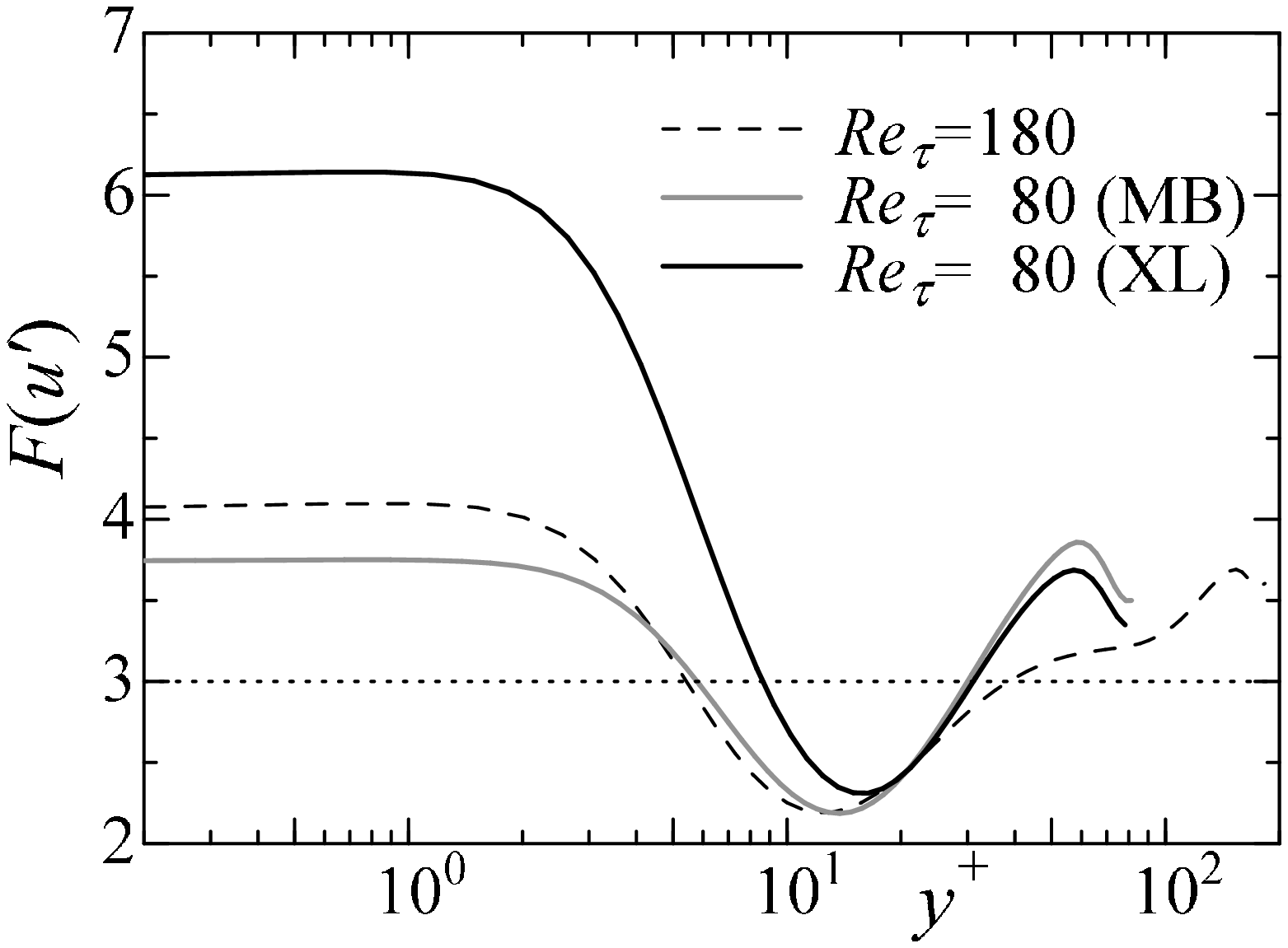}
   \end{center}
  \end{minipage}
  \hspace{-0.2em}
  \begin{minipage}[t]{0.49\textwidth}
   \begin{center}
    \epsfysize=45.0mm
    \epsfbox{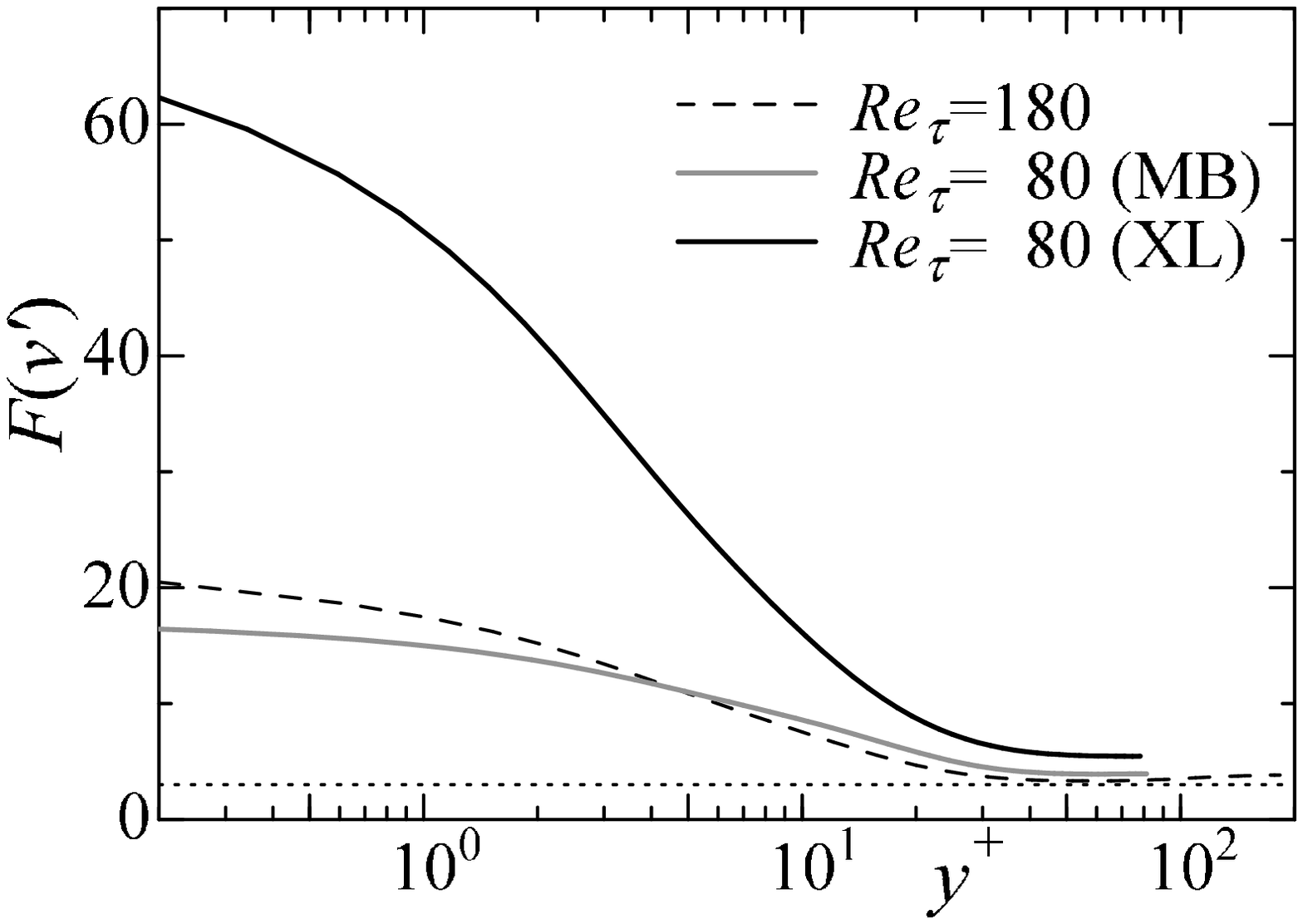}
   \end{center}
  \end{minipage}
 \end{tabular}\\
 \begin{center}
 \vspace{-8.0em}
 \hspace{+8.0em}(a) \hspace{14.8em} (b)\\
 \vspace{+6.5em}
 \caption{Flatness factor of velocity  fluctuation; 
          (a) streamwise component $F(u')$, (b) wall-normal component $F(v')$.
          }
 \label{fig:flat}
 \end{center}
\end{figure}

The flatness factor of the velocity fluctuation is shown in \fref{fig:flat}. 
For $Re_\tau=80$, comparison of the results with different box sizes indicates that the influence of box size on the flatness factor is significant. 
If the box size is extended large enough to capture the highly disordered turbulent region and weak-turbulent regions, intermittency of fluctuation is enhanced. 

\subsection{Turbulence structures: puff-like structure}

%%-------------------------------------- TIME SERIES OF U' & V'
\begin{figure}[tbp]
 \begin{center}
  \hspace{+6.2em}
  \epsfxsize=101.0mm
  \epsfbox{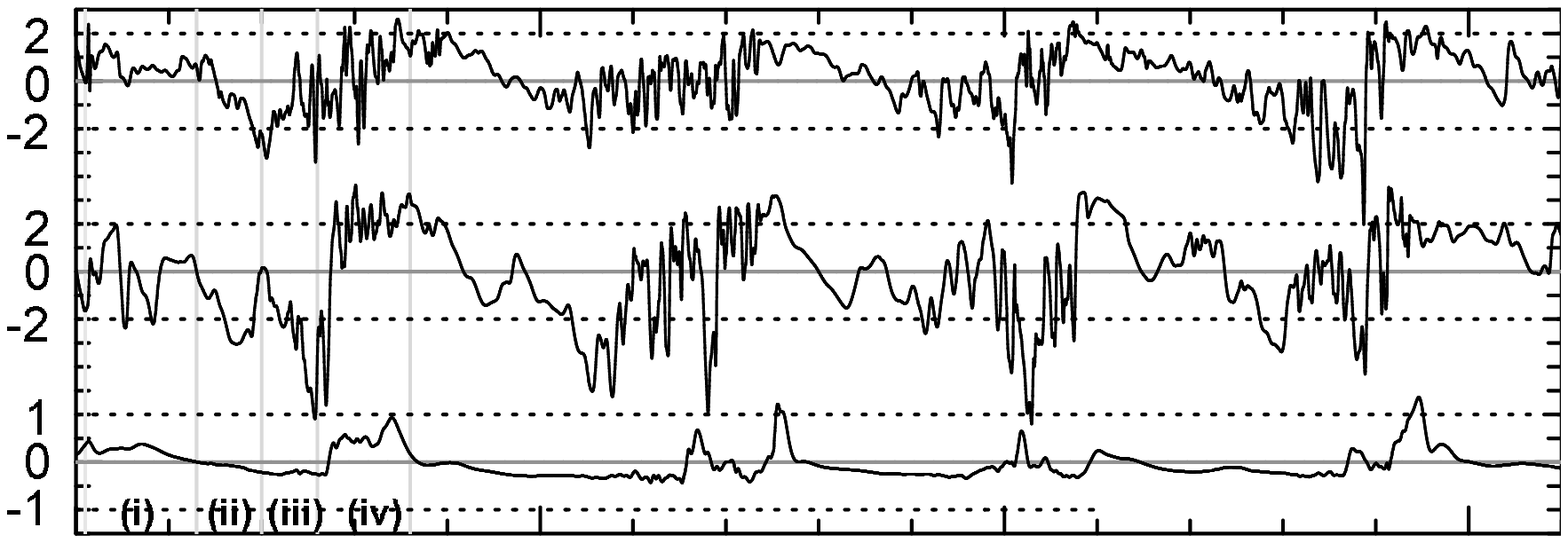} \\
  \vspace{-0.0em}
  \hspace{+5.2em}
  \epsfxsize=105.0mm
  \epsfbox{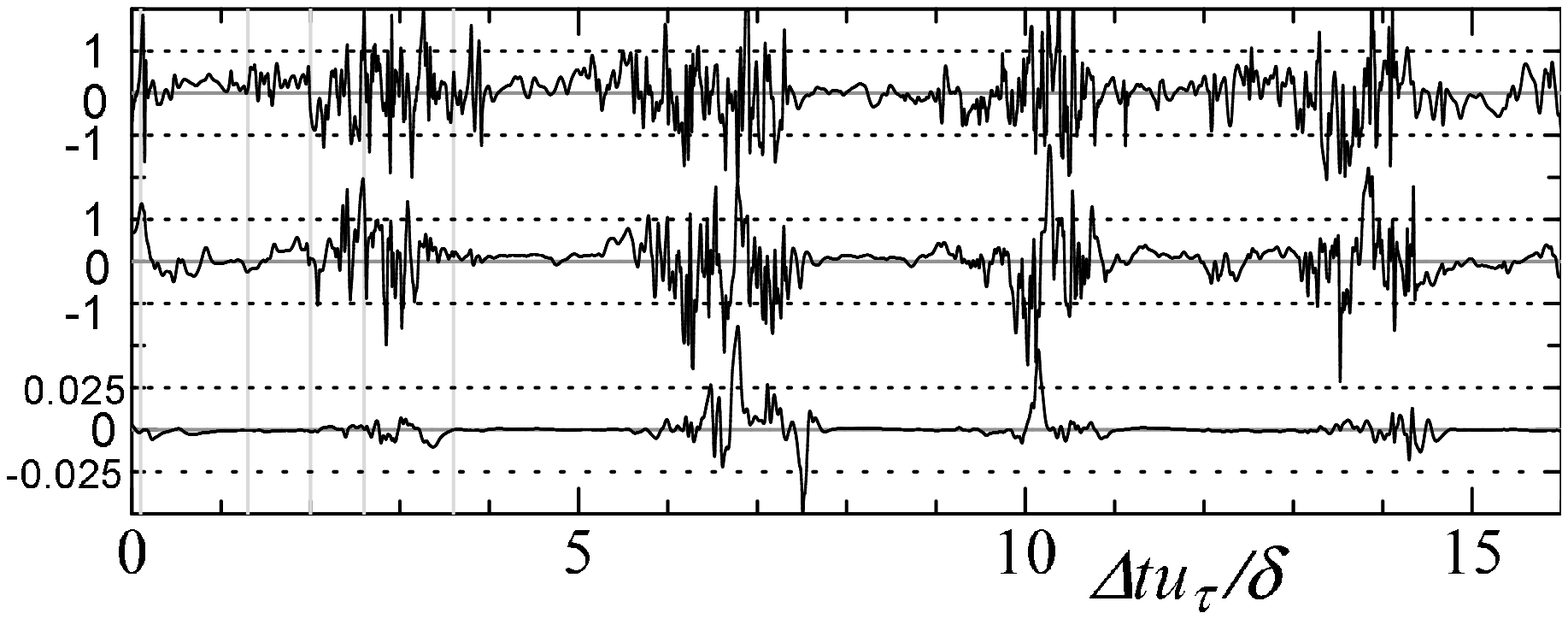} \\
 \vspace{-19.5em} (a) $y/\delta\approx$1~~ \hspace{26.8em}\, \\
 \vspace{+1.90em} (b) $y/\delta\approx$0.5 \hspace{26.8em}\, \\
 \vspace{+1.90em} (c) $y^+$=0.09           \hspace{26.8em}\, \\
 \vspace{+1.90em} (d) $y/\delta\approx$1~~ \hspace{26.8em}\, \\
 \vspace{+1.70em} (e) $y/\delta\approx$0.5 \hspace{26.8em}\, \\
 \vspace{+1.70em} (f) $y^+$=0.09           \hspace{26.8em}\, \\
 \vspace{+3.3em}
 \caption{Typical time trace of $u$- and $v$-fluctuations at $Re_\tau$=80~(XL). 
          The measurement points are at various wall-normal locations
          at the same position in the ($x,z$)-plane.
          (a)--(c), the streamwise fluctuation $u'^+$; 
          (d)--(f), the wall-normal fluctuation $v'^+$.
          }
 \label{fig:tmsrs}
 \end{center}
\end{figure}

Wygnanski and co-workers \cite{Wygnanski73, Wygnanski75} conducted a study of the structures and phenomena associated with transitional and turbulent \emph{pipe} flow in the range of $1,000<Re_{\rm m}<50,000$.
They identified two transitional flow states, the type observed being dependent on $Re_{\rm m}$.
For $2,000<Re_{\rm m}<2,700$, the transition structure was termed as `turbulent puff'; the second state was found at $Re_{\rm m}>3,500$ and consists of structures termed `turbulent slug'.
The puff and the slug are characterized by a distinct trailing edge over which the flow changes almost discontinuously from turbulent flow to laminar flow.
In the \emph{channel} flow of the present study, the computational results, shown in \fref{fig:tmsrs}, are strikingly similar to experimental velocity traces obtained by Wygnanski and Champagne \cite{Wygnanski73} for a \emph{pipe} flow.
\Fref{fig:tmsrs} shows a representative sample of $u'^+$ and $v'^+$ measured at several wall distances in the case of $Re_\tau=80$~(XL). 
It is observed that the sequence of events of the localized turbulence is advected past the measurement point.
It can be interpreted as: (i) quasi-laminar flow, (ii) a gradual reduction of $u$, (iii) a highly disordered turbulent region, (iv) a interface with quasi-laminar flow, again return to (i) quasi-laminar flow.
The intensities of both $u'$ and $v'$ are enhanced periodically with an interval of $\Delta t u_\tau/\delta \approx 3.5$, so the travel speed of the interface is estimated as $(L_x/\Delta t)^+ \approx 14.6$, which is close to the bulk mean velocity $u^+_{\rm m}\approx14.5$.
With respect to the present Reynolds number, i.e., $Re_{\rm m}$=2,320, the obtained structures of co-existing weak-turbulent region as discussed above are consistent with and remarkably similar to the `turbulent puffs' observed in a \emph{pipe} flow.

%%-------------------------------------- 2D T.P.C. 
\begin{figure}[tbp]
 \vspace{+0.0em}
 \begin{center}
  \hspace{+2.0em}
  \epsfxsize=60.0mm
  \epsfbox{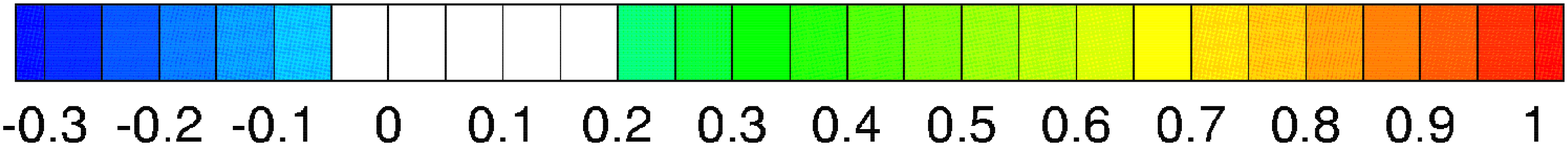} \\
  \vspace{-0.2em} (a) \hspace{30em} \, \\
  \vspace{-0.2em}
  \hspace{+2.5em}
  \epsfxsize=115.0mm
  \epsfbox{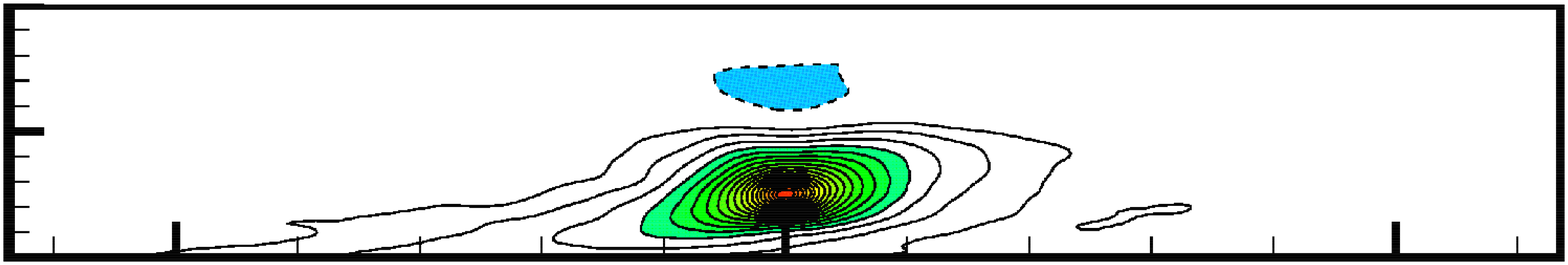} \\
  \vspace{+1.0em} (b) \hspace{30em} \, \\
  \vspace{-0.2em}
  \hspace{+2.5em}
  \epsfxsize=115.0mm
  \epsfbox{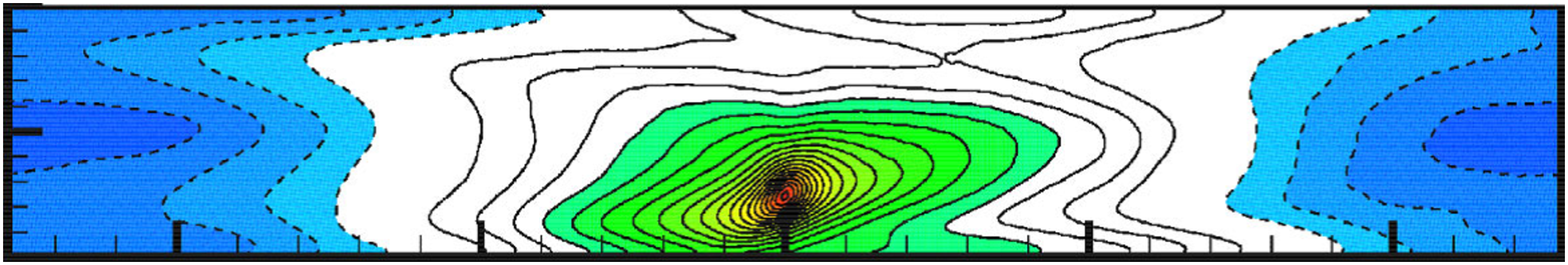} \\
 \vspace{-15.0em} $R_{uu}(\Delta x,y)$ \hspace{18em}\, \\
 \vspace{+1.0em}  2         \hspace{27.5em}\, \\
 \vspace{-0.0em}  $y/\delta$\hspace{30em}\, \\
 \vspace{+0.0em}  1         \hspace{27.5em}\, \\
 \vspace{+1.0em}  0         \hspace{27.5em}\, \\
 \vspace{-0.5em}  \hspace{5.5em} -5 \hspace{10.5em} 0 \hspace{10.5em} 5 \hspace{2.5em} \,\\
 \vspace{-1.0em}  \hspace{12.0em} $\Delta x/\delta$ \\
 \vspace{+0.5em}  2         \hspace{27.5em}\, \\
 \vspace{-0.0em}  $y/\delta$\hspace{30em}\, \\
 \vspace{+0.0em}  1         \hspace{27.5em}\, \\
 \vspace{+1.0em}  0         \hspace{27.5em}\, \\
 \vspace{-0.4em}  \hspace{5.0em} -20 \hspace{10.4em} 0 \hspace{10.1em} 20 \hspace{2.5em} \,\\
 \vspace{-1.0em}  \hspace{12.0em} $\Delta x/\delta$ \\
 \vspace{+1.0em}
 \caption{Contours of two-dimensional two-point correlation coefficient $R_{uu}$ 
          in the ($x,y$)-plane; 
          the reference point is at the mid-height $y_{\textrm{\tiny{ref}}}=0.5\delta$.
          The direction of the mean flow is from left to right.
          (a) $Re_\tau=180$; (b) $Re_\tau=80$ (XL).
          \full, positive correlation; \dashed, negative correlation; 
          the line of $R_{uu}=0$ is not shown here.
          Contour interval $\Delta R_{uu}$:0.05.
          }
 \label{fig:corr2d}
 \end{center}
\end{figure}

The shape of structure in the wall-normal direction can be clearly illustrated by the two-dimensional correlation for velocity in the ($x,y$)-plane.
The fixed point is $y_\textrm{\tiny{ref}}^+=5$.
The correlation coefficient is defined as 
\begin{equation}
 R_{uu}(\Delta x,y)=\frac{\overline{u'(x,y_\textrm{\tiny{ref}},z)u'(x+\Delta x,y,z)}}
                          {u'_\textrm{\tiny{rms}}(y_\textrm{\tiny{ref}})u'_\textrm{\tiny{rms}}(y)}.
 \label{eq:corr2d}
\end{equation}
\noindent{} While a region of negative correlation appears on the side of the other wall for $Re_\tau=180$ (\fref{fig:corr2d}(a)), a negative region appears in a wide region of $|\Delta x|>13\delta$ for $Re_\tau=80$~(XL).
This indicates that a puff-like structure with a large streamwise wavelength of 51.2$\delta$ fills 
the entire channel width in the case of $Re_\tau=80$~(XL).

%%------------------------------------------------------------- ANIMATION
\begin{figure}[t!]
  \begin{center}
   (a) $\Delta t u_\tau/\delta = 0$ \\
   \hspace{1em}
   \epsfxsize=110.0mm
   \epsfbox{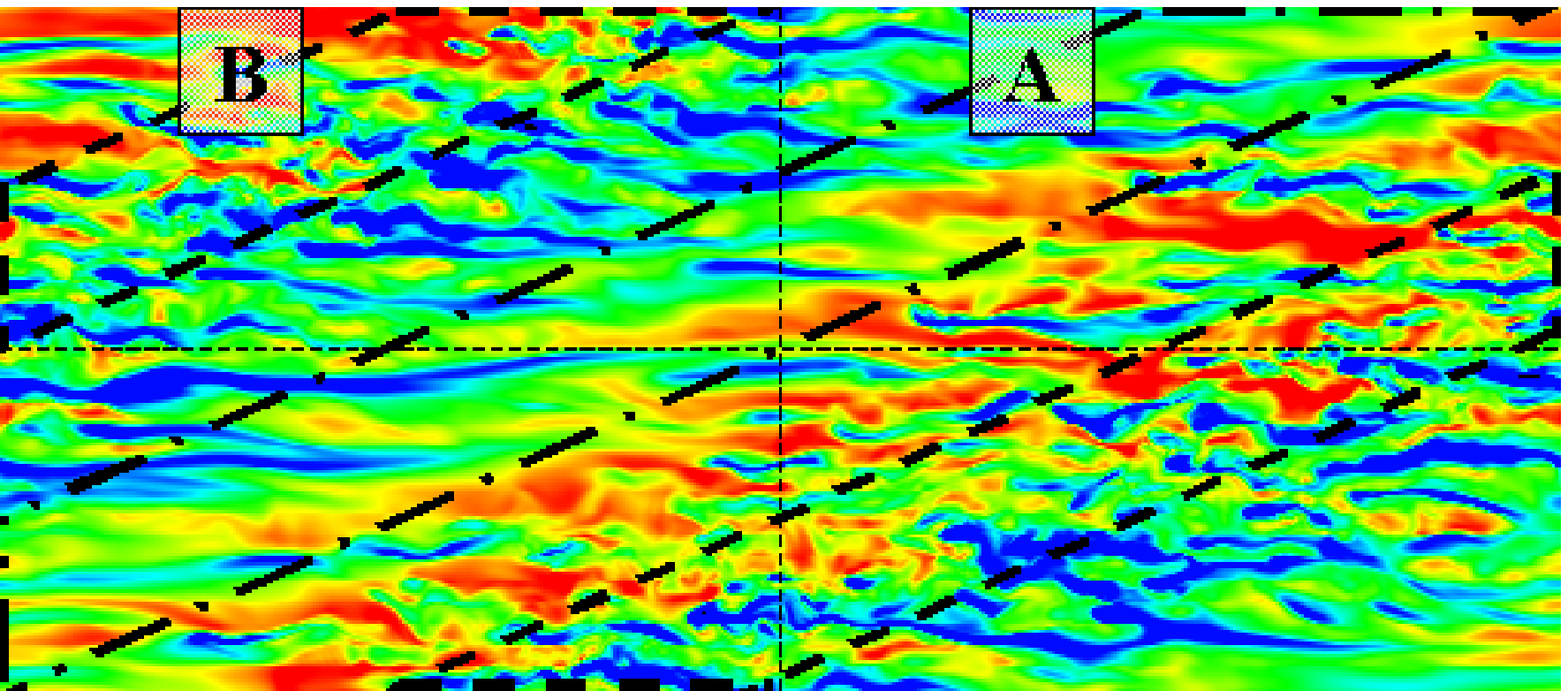} \\
  \end{center}
  \vspace{-14.0em}  \hspace{1em}22.5 \\
  \vspace{+3.5em}\\ \hspace{1em}$z/\delta$ \\
  \vspace{+4.8em}\\ \hspace{2.5em}0 \hspace{13.0em} $x/\delta$ \hspace{11.2em} 51.2\\
  \vspace{-0.8em}\\
 \begin{tabular}{cc}
 \hspace{-0.5em}
 \begin{minipage}[t]{0.49\textwidth}
  \begin{center}
   (b) $\Delta t u_\tau/\delta = 0.36$ \\
   \epsfxsize=60.0mm
   \epsfbox{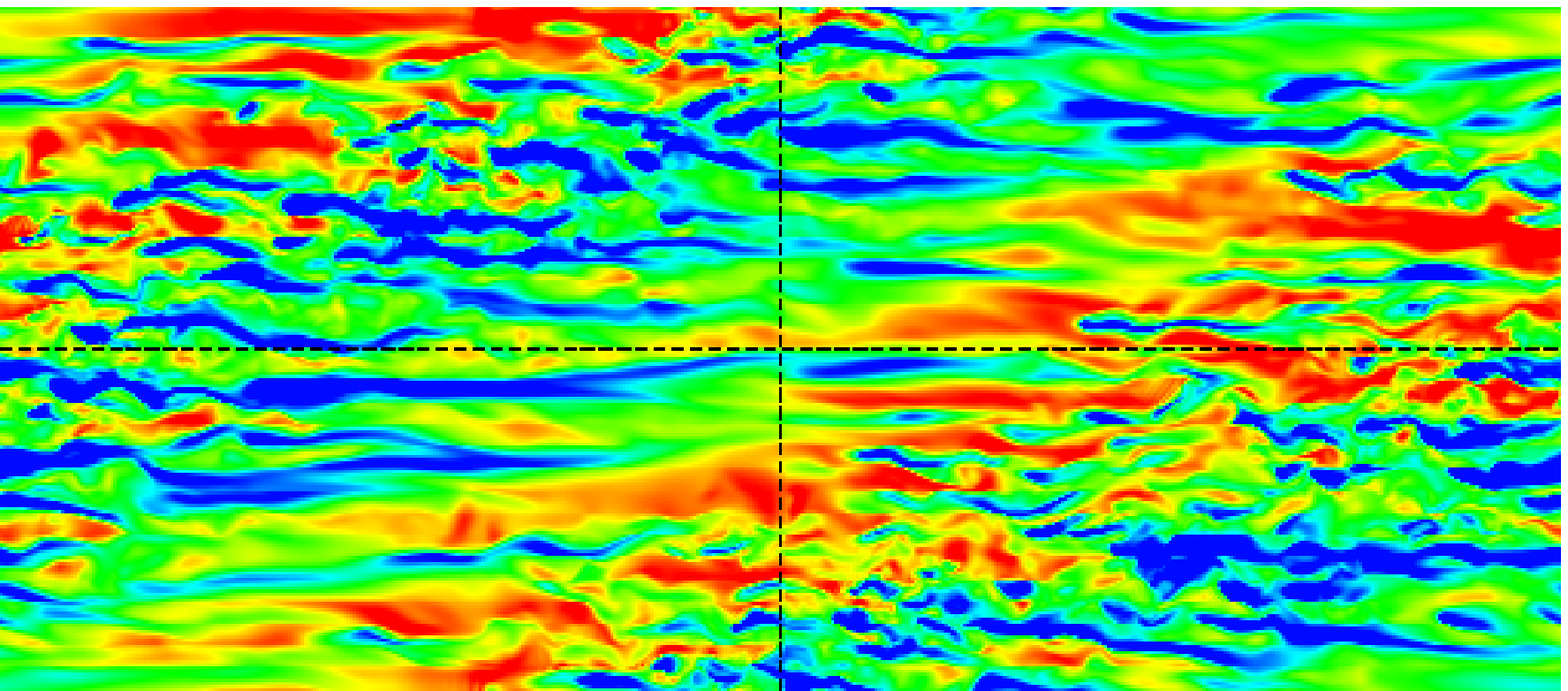}
  \end{center}
 \end{minipage}
 \begin{minipage}[t]{0.49\textwidth}
  \begin{center}
   (f) $\Delta t u_\tau/\delta = 1.80$ \\
   \epsfxsize=60.0mm
   \epsfbox{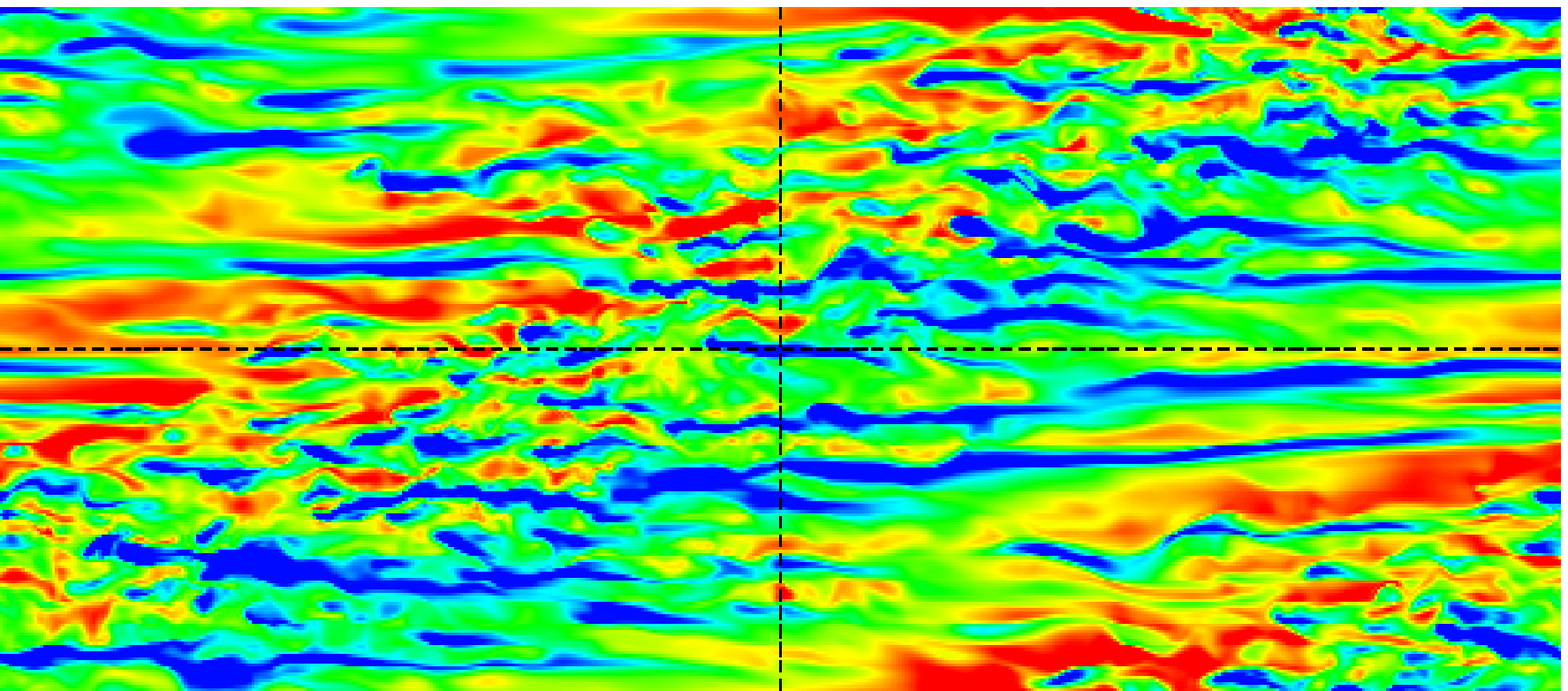}
  \end{center}
 \end{minipage}
 \end{tabular}
 \begin{tabular}{cc}
 \hspace{-0.5em}
 \begin{minipage}[t]{0.49\textwidth}
  \begin{center}
   (c) $\Delta t u_\tau/\delta = 0.72$ \\
   \epsfxsize=60.0mm
   \epsfbox{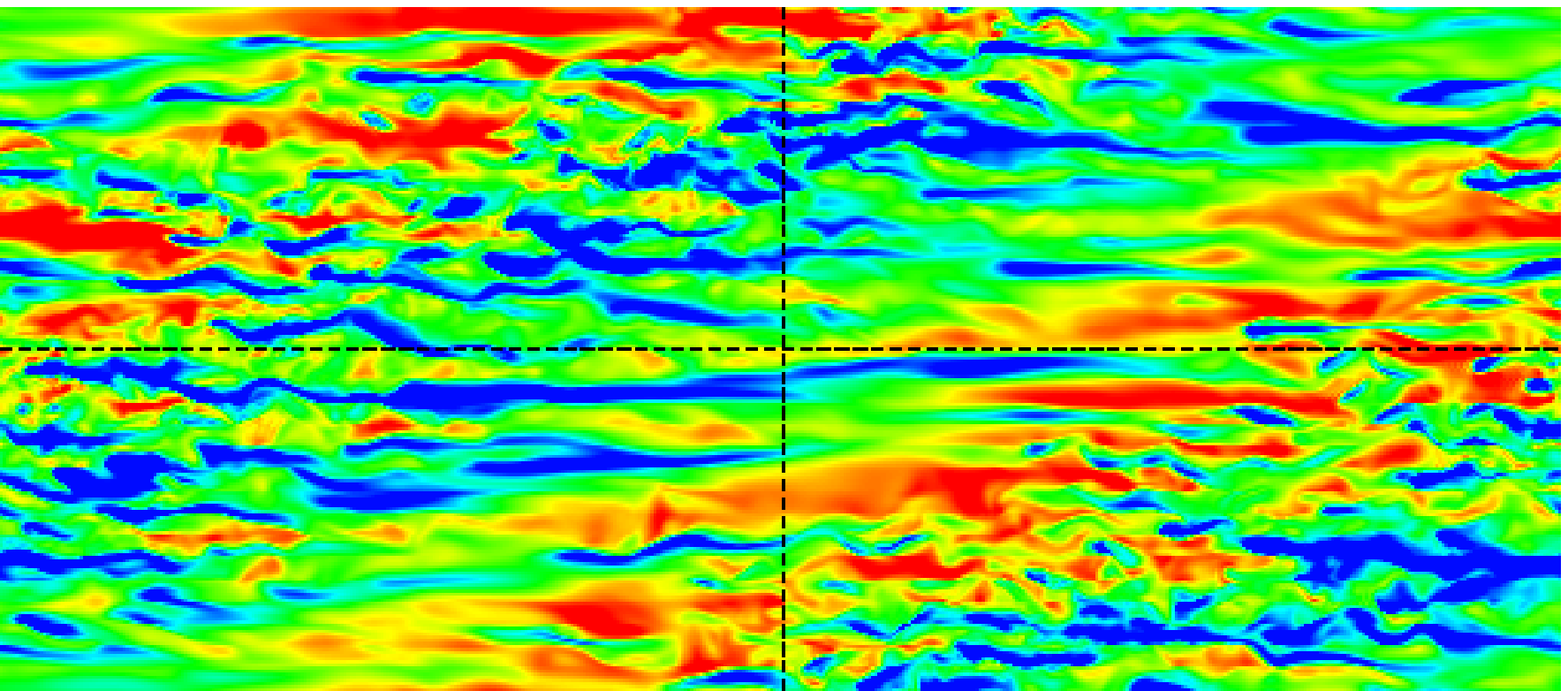}
  \end{center}
 \end{minipage}
 \begin{minipage}[t]{0.49\textwidth}
  \begin{center}
   (g) $\Delta t u_\tau/\delta = 2.16$ \\
   \epsfxsize=60.0mm
   \epsfbox{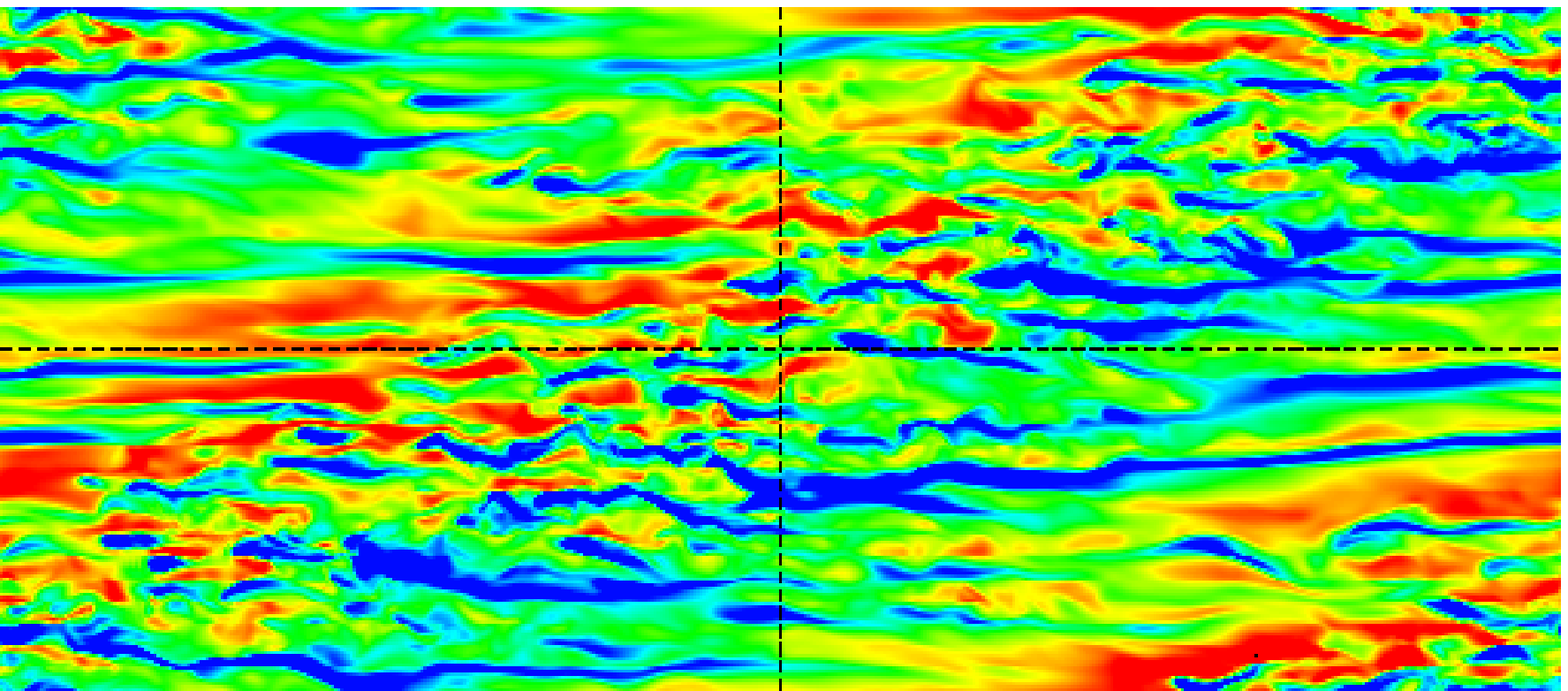}
  \end{center}
 \end{minipage}
 \end{tabular}
 \begin{tabular}{cc}
 \hspace{-0.5em}
 \begin{minipage}[t]{0.49\textwidth}
  \begin{center}
   (d) $\Delta t u_\tau/\delta = 1.08$ \\
   \epsfxsize=60.0mm
   \epsfbox{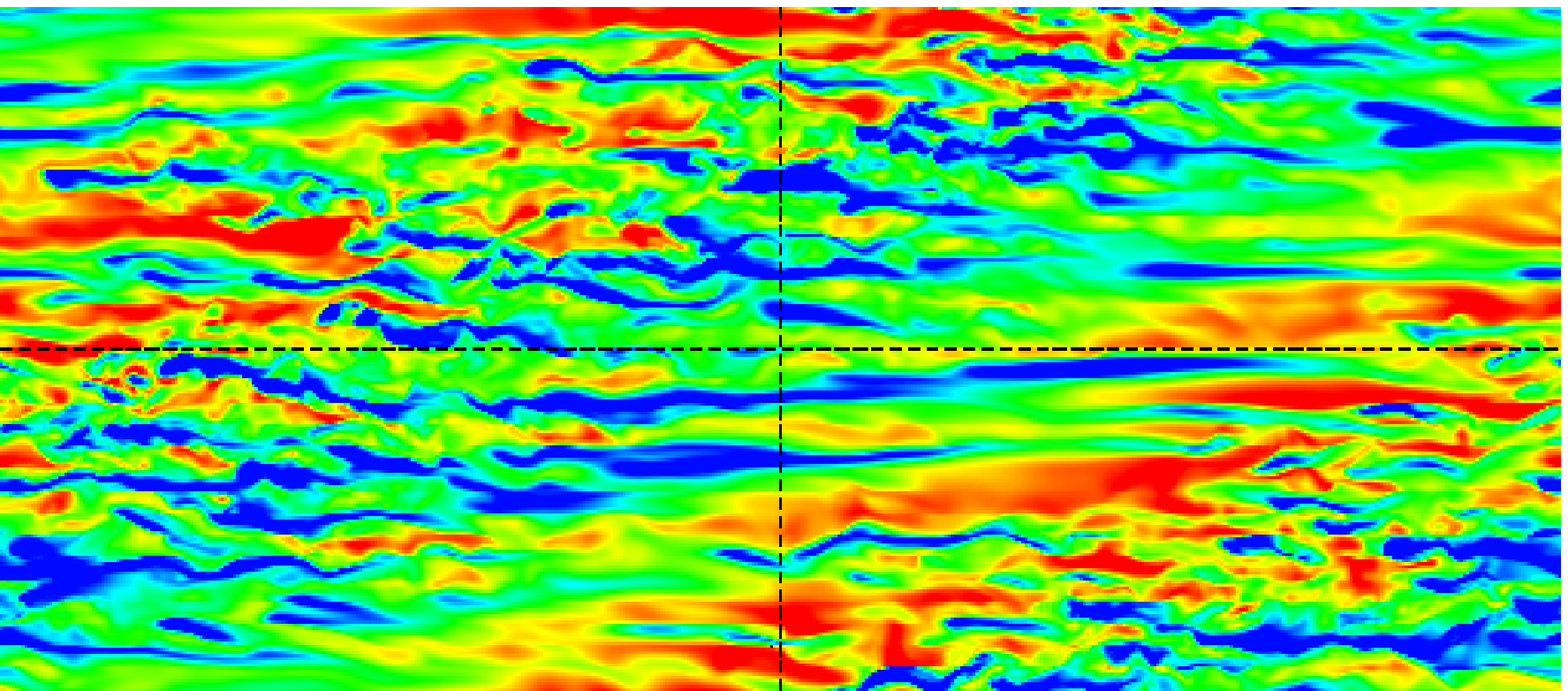}
  \end{center}
 \end{minipage}
 \begin{minipage}[t]{0.49\textwidth}
  \begin{center}
   (h) $\Delta t u_\tau/\delta = 2.52$ \\
   \epsfxsize=60.0mm
   \epsfbox{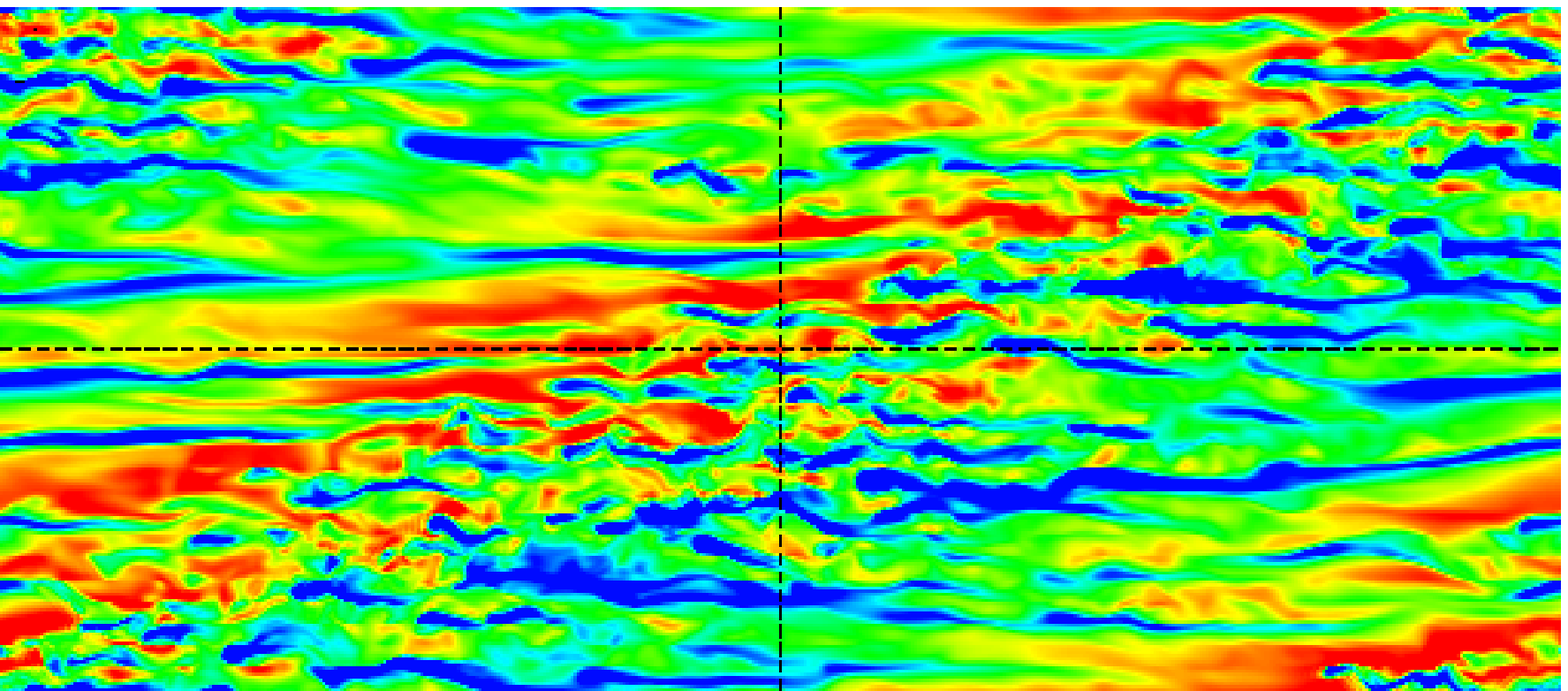}
  \end{center}
 \end{minipage}
 \end{tabular}
 \begin{tabular}{cc}
 \hspace{-0.5em}
 \begin{minipage}[t]{0.49\textwidth}
  \begin{center}
   (e) $\Delta t u_\tau/\delta = 1.44$ \\
   \epsfxsize=60.0mm
   \epsfbox{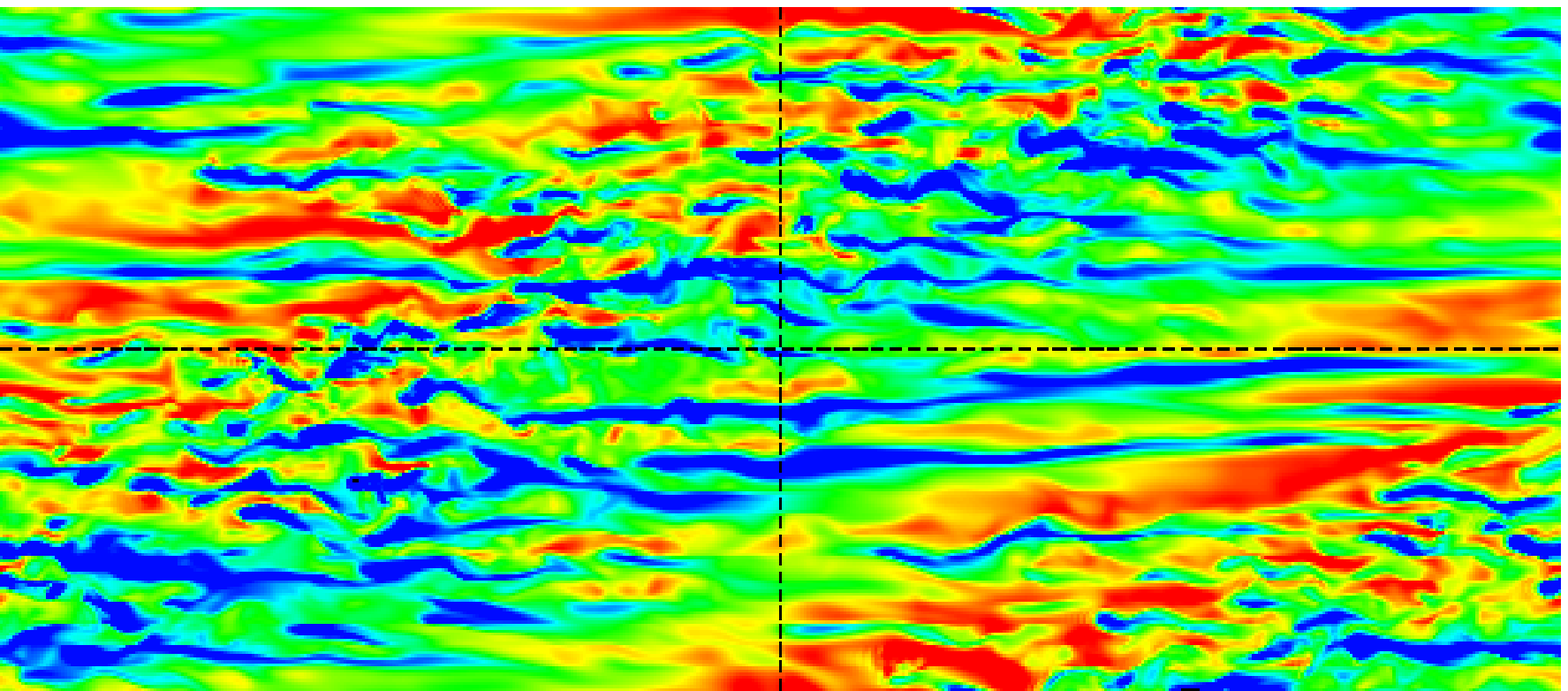}
  \end{center}
 \end{minipage}
 \begin{minipage}[t]{0.49\textwidth}
  \begin{center}
  \vspace{1em}
   ${u'}^+$\\
   \vspace{0.3em}
   \epsfxsize=60.0mm
   \epsfbox{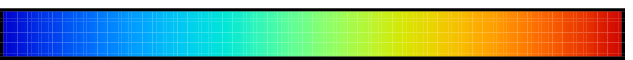} \\
   $-3.0$ \hspace{4.5em} 0.0 \hspace{4.5em} $+3.0$
  \end{center}
 \end{minipage}
 \end{tabular}
 \vspace{-0.2em}
 \caption{Contour of streamwise velocity fluctuation ${u'}^+$ 
          for a series of consecutive times,
          in an ($x,z$)-plane at $y/\delta \approx 0.5$ for $Re_\tau=80$ (XL).
          All of the contours (a)--(f) shows the whole ($x,z$)-plane of 
          ($L_x \times L_z$)=($51.2\delta \times 22.5\delta$).
          The direction of the mean flow is from left to right.
          }
 \label{fig:anim}
 \vspace{+0.5em}
\end{figure}

A typical well-developed structure at $Re_\tau=80$~(XL) is shown as a sequence of flow visualizations in \fref{fig:anim}.
The puff-like structure, shown in \fref{fig:anim}(a), consists of a quasi-laminar state (region marked A) and a highly disordered turbulence (region B). 
The wavelength of the near-wall streaks is much shorter than that of the spatial distribution of large-scale fluctuations induced by the puff-like structure, since the former is scaled in wall units, i.e., $\lambda_z^+ =100$, and the latter is almost equal to the box size as shown in \sref{sec:pmes}. 
The actual flow field is a superposition of these two kinds of structures.
In consequence, the regions where either high- or low-speed streak is dominant are discriminable, as seen in figures~\ref{fig:instfld}(b) and \ref{fig:anim}.

Figures \ref{fig:anim}(b)-(h) indicate that the puff-like structure is equilibrium and self-sustained.
Moreover, the both regions (A and B) propagate with the streamwise velocity same with the bulk velocity and are inclined at an angle of $24^\circ$ with respect to the streamwise direction.
(This angle is easily determined by the aspect ratio of the horizontal domain size --- calculated from $\tan^{-1}(L_z/L_x)$ --- in the case of the present result.) 
Therefore, the puff-like structures are spatially distributed not only in streamwise but also spanwise direction, while the puff of a \emph{pipe} flow is homogeneous in azimuthal and intermittent only in the streamwise direction.
The angle of the oblique structure can be affected by the aspect ratio and the dimension of the computation domain. 
As discussed in the previous section, the similar oblique structure has been observed in an experiment of the plane Couette flow \cite{Prigent03}. 
One may regard the inclination of the puff-like structure as essential in the transitional channel flows.
Note that the present results, such as visualized flow fields, two-point correlation coefficients (\fref{fig:tpc}) and  energy spectra (\fref{fig:pmes}), show signs of being constrained by the periodicity of the boundary, even when the box size was extended to the XL. 
The numerical box requires to be enlarged further to allow us to neglect the periodicity of the computational domain. 

\section{Conclusions}

In the present study, we performed DNS of the turbulent channel flow with larger computational boxes and investigated the turbulence statistics with respect to low-Reynolds-number effect.
The Reynolds number was decreased down to $Re_\tau=60$ to study the characteristics of the transitional channel flow and to estimate the critical Reynolds number at which a laminarization occurs.
Based on the results of the computations we can draw the following conclusions:

\begin{enumerate}
\item Turbulent/transitional state can be sustained in the computational box size of $25.6\delta \times 2\delta \times 12.8\delta$ in the range of $Re_\tau\ge64$.

\item For $Re_{\rm m}<3,000$, $C_f$ tends to be smaller than the empirical correlation and decreases with decreasing Reynolds numbers for $Re_{\rm m} < 2,000$.

\item The Reynolds-number dependence of the mean velocity profile is significant in the outer region when scaled with the wall units. The turbulence statistics indicate that the anisotropy of turbulence is enhanced at low Reynolds numbers.

\item For $Re_\tau=80$ ($Re_{\rm m}=2,320$), the periodic localized turbulence is observed with using the largest box of $51.2\delta\times2\delta\times22.5\delta$ and very similar to a `turbulent puff' observed in a transitional pipe flow.

\item The significant influence of the captured puff-like structures exist on the turbulence statistics, such as a mean velocity, turbulence intensities and vorticity fluctuations.

\item The equilibrium puff-like structures observed in the channel flow inclines against the streamwise direction. The propagation velocity of the puff-like structure is approximately equal to the bulk mean velocity.
\end{enumerate}

\section*{Acknowledgements}

The present study is entrusted from Ministry of Education, Culture, Sports, Science and Technology of Japan.
The computations were performed with the use of supercomputing resources at Information Synergy Center of Tohoku University, and also VPP5000/64 at Computing and Communications Center of Kyushu University.
The author would like to thank Dr. Kaoru Iwamoto for fruitful discussions and careful reading of the manuscript.

\vspace{1em}

This paper is a revised and expanded version of a paper entitled ``DNS of turbulent channel flow at very low Reynolds numbers'', presented by T. Tsukahara, Y. Seki, H. Kawamura, and D. Tochio, at the 4th Int. Symp. on Turbulence and Shear Flow Phenomena, Williamsburg, VA, USA, Jun. 27--29 (2005), pp. 935--940.

\label{lastpage}


\begin{thebibliography}{99}

\bibitem{Narashimha79}
Narasimha, R. and Sreenivasan, K. R., 1979, 
Relaminarization of Fluid Flows. 
{\itshape Advances in Applied Mechanics}, \textbf{19}, 221--309.

\bibitem{Kim87}
Kim, J. Moin, P. and Moser, R. D., 1987, 
Turbulence statistics in fully developed turbulent channel flow at low Reynolds number. 
{\itshape Journal of Fluid Mechanics}, \textbf{177}, 133--166.

\bibitem{Kuroda89}
Kuroda, A., Kasagi, N. and Hirata, M., 1989, 
A direct numerical simulation of the fully developed turbulent channel flow 
at a very low Reynolds number. 
Proceedings of the 3rd International Symposium on Computational Fluid Dynamics,
Nagoya, Japan, August, pp.1174--1179.

\bibitem{Kawamura98}
Kawamura, H., Ohsaka, K., Abe, H. and Yamamoto, K., 1998, 
DNS of turbulent heat transfer in channel flow with low to medium-high Prandtl number fluid. 
{\itshape International Journal of Heat and Fluid Flow}, \textbf{19}, 482--491.

\bibitem{Abe01}
Abe, H., Kawamura, H. and Matsuo, Y., 2001, 
Direct numerical simulation of a fully developed turbulent channel flow 
with respect to the Reynolds number dependence. 
{\itshape Transactions of the ASME}. I: {\itshape Journal of Fluids Engineering}, 
\textbf{123}, 382--393.

\bibitem{Abe04a}
Abe, H., Kawamura, H. and Choi, H., 2004, 
Very large-scale structures and their effects 
on the wall shear-stress fluctuations in a turbulent channel flow up to $Re_\tau=640$. 
{\itshape Transactions of the ASME}. I: {\itshape Journal of Fluids Engineering}, 
\textbf{126}, 835--843.

\bibitem{Abe04b}
Abe, H., Kawamura, H. and Matsuo, Y., 2004, 
Surface heat-flux fluctuations in a turbulent channel flow up to $Re_\tau$=1020 
with $Pr$=0.025 and 0.71. 
{\itshape International Journal of Heat and Fluid Flow}, \textbf{25}, 404--419.

\bibitem{Bewley01}
Bewley, T. R., Moin, P. and Temam, R., 2001, 
DNS-based predictive control of turbulence: an optimal benchmark for feedback algorithms. 
{\itshape Journal of Fluid Mechanics}, \textbf{447}, 179--225.

\bibitem{Chang02}
Chang, Y., Collis, S. S. and Ramakrishman, S., 2002, 
Viscous effects in control of near-wall turbulence. 
{\itshape Physics of Fluids}, \textbf{14}, 4069--4080.

\bibitem{Hogberg03}
H\"ogberg, M., Bewley, T. R. and Henningson, D. S., 2003, 
Relaminarization of $Re_\tau$=100 turbulence using gain scheduling 
and linear state-feedback control. 
{\itshape Physics of Fluids}, \textbf{15}, 3572--3575.

\bibitem{Iwamoto02}
Iwamoto, K., Suzuki, Y. and Kasagi, N., 2002, 
Reynolds number effect on wall turbulence : toward effective feedback control. 
{\itshape International Journal of Heat and Fluid Flow}, \textbf{23}, 678--689.

\bibitem{Iida98}
Iida, O. and Nagano, Y., 1998, 
The relaminarization mechanisms of turbulent channel flow at low Reynolds numbers. 
{\itshape Flow, Turbulence and Combustion}, \textbf{60}, 193--213.

\bibitem{Laufer51}
Laufer, J., 1951, 
Investigation of turbulent flow in a two-dimensional channel. 
NACA Report, 1053, 1247--1266.

\bibitem{Patel68}
Patel, V.C. and Head, M.R., 1968, 
Reversion of turbulent to laminar flow. 
{\itshape Journal of Fluid Mechanics}, \textbf{34}, 371--392.

\bibitem{Patel69}
Patel, V.C. and Head, M.R., 1969, 
Some observations on skin friction and velocity profiles 
in fully developed pipe and channel flows. 
{\itshape Journal of Fluid Mechanics}, \textbf{38}, 181--201.

\bibitem{Eckelmann74}
Eckelmann, H., 1974, 
The structure of the viscous sublayer and the adjacent wall region in a turbulent channel flow. 
{\itshape Journal of Fluid Mechanics}, \textbf{65}, 439--459.

\bibitem{Kreplin79}
Kreplin, H. P. and Eckelmann, H., 1979, 
Behavior of the three fluctuating velocity components in the wall region of 
a turbulent channel flow. 
{\itshape Physics of Fluids}, \textbf{22}, 1233--1239.

\bibitem{Niederschulte90}
Niederschulte, M. A., Adrian, R. J. and Hanratty, T. J., 1990, 
Measurements of turbulent flow in a channel at low Reynolds numbers. 
{\itshape Experiments in Fluids}, \textbf{9}, 222--230.

\bibitem{Durst95}
Durst, F. and Kikura, H., 1995, 
Low Reynolds number effects on a fully developed turbulent channel flow. 
Proceedings of the 10th Symposium on Turbulent shear flows,
Pennsylvania, USA, 14--16 August, P2-25--30.

\bibitem{Davies28}
Davies, S.J. and White, C. M., 1928, 
An experimental study of the flow of water in pipes of rectangular section. 
{\itshape Proceeding of Royal Society London A}, \textbf{119}, 92--107.

\bibitem{Kao70}
Kao, T. W. and Park, C., 1970, 
Experimental investigations of the stability of channel flows. 
Part 1. Flow of a single liquid in a rectangular channel. 
{\itshape Journal of Fluid Mechanics}, \textbf{43}, 145--164.

\bibitem{Nishioka75}
Nishioka, M., Iida, S. and Ichikawa, Y., 1975, 
An experimental investigation of the stability of plane Poiseuille flow. 
{\itshape Journal of Fluid Mechanics}, \textbf{72}, 731--751.

\bibitem{Orszag71}
Orszag, S. A., 1971, 
Accurate solution of the Orr Sommerfeld stability equation. 
{\itshape Journal of Fluid Mechanics}, \textbf{50}, 689--703.

\bibitem{Orszag80a}
Orszag, S. A. and Kells, L. C., 1980, 
Transition to turbulence in plane Poiseuille flow and plane Couette flow. 
{\itshape Journal of Fluid Mechanics}, \textbf{96}, 159--205.

\bibitem{Orszag80b}
Orszag, S. A. and Patera, A. T., 1980, 
Subcritical transition to turbulence in plane channel flows. 
{\itshape Physical Review Letters}, \textbf{45}, 989--993.

\bibitem{Kleiser91}
Kleiser, L. and Zang, T. A., 1991, 
Numerical simulation of transition in wall-bounded shear flows. 
{\itshape Annual Review of Fluid Mechanics}, \textbf{23}, 495--537.

\bibitem{Butler92}
Butler, K. M. and Farrell, B. F., 1992
Three-dimensional optimal perturbations in viscous shear flow. 
{\itshape Physics of Fluids}, A\textbf{4}, 1637--1650.

\bibitem{Reddy93}
Reddy, S. C. and Henningson, D. S., 1993, 
Energy growth in viscous channel flows. 
{\itshape Journal of Fluid Mechanics}, \textbf{252}, 209--238.

\bibitem{Jimenez98}
Jim\'enez, J., 1998, 
The largest scales of turbulent wall flows. 
{\itshape Center for Turbulence Research Annual Research Briefs}, 137--154.

\bibitem{Liu01}
Liu Z., Adrian R. J. and Hanratty T. J., 2001, 
Large-scale modes of turbulent channel flow: transport and structure. 
{\itshape Journal of Fluid Mechanics}, \textbf{448}, 53--80.

\bibitem{Dean78}
Dean, R. D., 1978, 
Reynolds number dependence of skin friction and other bulk flow variables 
in two-dimensional rectangular duct flow. 
{\itshape Transactions of the ASME}. I: {\itshape Journal of Fluids Engineering}, 
\textbf{100}, 215--222.

\bibitem{Carlson82}
Carlson, D. R.,Widnall, S. E. and Peeters, M. F., 1982, 
A flow-visualization study of transition in plane Poiseuille flow. 
{\itshape Journal of Fluid Mechanics}, \textbf{121}, 487--505.

\bibitem{Thomas53}
Thomas, L. H., 1953, 
The stability of plane Poiseuille flow. 
{\itshape Physical Review}, \textbf{91}, 780--783.

\bibitem{Antonia92}
Antonia, R. A., Teitel, M., Kim, J. and Browne, L. W. B., 1992, 
Low-Reynolds-number effects in a fully developed turbulent channel flow. 
{\itshape Journal of Fluid Mechanics}, \textbf{236}, 579--605.

\bibitem{Antonia94}
Antonia, R. A. and Kim, J., 1994, 
Low-Reynolds-number effects on near-wall turbulence. 
{\itshape Journal of Fluid Mechanics}, \textbf{276}, 61--80.

\bibitem{Sahay99}
Sahay, A. and Sreenivasan, K. R., 1999 
The wall-normal position in pipe and channel flows at which viscous 
and turbulent shear stresses are equal. 
{\itshape Physics of Fluids}, \textbf{11}, 3186--3188.

\bibitem{Laadhari02}
Laadhari, F., 2002 
On the evolution of maximum turbulent kinetic energy production in a channel flow. 
{\itshape Physics of Fluids}, \textbf{14}, L65--L68.

\bibitem{Wei89}
Wei, T. and Willmarth, W. W., 1989, 
Reynolds-number effects on the structure of a turbulent channel flow. 
{\itshape Journal of Fluid Mechanics}, \textbf{204}, 57--95.

\bibitem{Perry86}
Perry, A. E., Henbest, S. and Chong M. S., 1986, 
A theoretical and experimental study of wall turbulence. 
{\itshape Journal of Fluid Mechanics}, \textbf{165}, 163--199.

\bibitem{Wygnanski73}
Wygnanski, I.J. and Champagne, F.H., 1973, 
On transition in a pipe. Part 1. The origin of puffs and slugs and the flow in a turbulent slug. 
{\itshape Journal of Fluid Mechanics}, \textbf{59}, 281--335.

\bibitem{Wygnanski75}
Wygnanski, I., Sokolov, M. and Friedman, D., 1975, 
On transition in a pipe. Part 2. The equilibrium puff. 
{\itshape Journal of Fluid Mechanics}, \textbf{69}, 283--304.

\bibitem{Prigent03}
Prigent, A., Gregoire, G., Chate, H., Dauchot, O. and van Saarloos, W., 2002, 
Large-scale finite-wavelength modulation within turbulent shear flows. 
{\itshape Physical Review Letters}, \textbf{89}, 014501.

\end{thebibliography}
\end{document}